\documentclass[twocolumn]{openjournal}

\usepackage[utf8]{inputenc}
\usepackage[T1]{fontenc}
\usepackage{float}
\usepackage{ulem}
\usepackage{graphicx}	
\usepackage{amsmath}	
\usepackage{longtable}
\usepackage{comment}
\usepackage[dvipsnames,svgnames]{xcolor}
\usepackage{color}
\usepackage{hyperref}
\hypersetup{colorlinks=true,
	urlcolor=blue,  
	linkcolor=blue,  
	citecolor=blue, 
    menucolor=blue, 
    urlcolor=blue}  

\def\msun{\, M_{\odot}}

\def\simlt{\lower.5ex\hbox{$\; \buildrel < \over \sim \;$}}
\def\simgt{\lower.5ex\hbox{$\; \buildrel > \over \sim \;$}}

\newcommand{\astrid}{\texttt{ASTRID}}
\newcommand{\brahma}{\texttt{BRAHMA}}
\newcommand{\ambra}{\texttt{AMBRA}}
\def\AGNstoch{0.5}
\newcommand{\redcolorLabel}{``Red Color''} 
\newcommand{\bluecolorLabel}{``Blue Color''} 
\newcommand{\RedColorText}{The 2D histogram shows the overall \ambra\ population. It is colored by the mean \redcolorLabel\ in each cell, where \redcolorLabel\ is F200W-F356W for $z=5$, 6 sources, and F277W-F444W for $z=7$, 8 sources}

\newcommand{\icontext}{The LRDs identified in \ambra\ are marked with red circles} 
\newcommand{\shadebetweentext}{The shaded region indicates the Poisson uncertainty of the results. As the gas-enshrouded AGN results are an upper limit, we shade between the upper Poisson uncertainty of the gas-enshrouded results, and the lower Poisson uncertainty of the standard AGN results where applicable}
\newcommand{\tracktext}{We show the redshift evolution of the two sources shown in figure \ref{fig:histories_w_images} as green and gold tracks. The direction of redshift evolution is indicated by the arrows along the tracks, and the snapshots where they are observed as LRDs are shown with circles of the matching color}
\newcommand{\meanlines}{We include the mean of LRD and overall populations as the solid and dashed violet lines respectively}

\begin{document}

\title{The lifetimes and properties of Little Red Dots in the AMBRA simulation}

\author{Patrick LaChance$^{1,*}$}
\author{Yihao Zhou$^{1}$}
\author{Aklant Kumar Bhowmick$^{2,3,4}$}
\author{Rupert A.C. Croft$^{1}$}
\author{Tiziana Di Matteo$^{1}$}
\author{Laura Blecha$^{5}$}
\author{Fabio Pacucci$^{6}$}
\author{Paul Torrey$^{2,3,4}$}
\author{Simeon Bird$^{7}$}


\thanks{$^*$E-mail: plachance@cmu.edu}
\affiliation{$^{1}$ McWilliams Center for Cosmology and Astrophysics, Department of Physics, Carnegie Mellon University, Pittsburgh, PA 15213, USA}
\affiliation{$^{2}$ University of Virginia, 530 McCormick Rd, Charlottesville, VA 22904, USA}
\affiliation{$^{3}$ Virginia Institute for Theoretical Astronomy, University of Virginia, Charlottesville, VA 22904, USA }
\affiliation{$^{4}$ The NSF-Simons AI Institute for Cosmic Origins, USA }
\affiliation{$^{5}$ Department of Physics, University of Florida, Gainesville, FL 32611, USA}
\affiliation{$^{6}$ Center for Astrophysics $\vert$ Harvard \& Smithsonian, 60 Garden St, Cambridge, MA 02138, USA}
\affiliation{$^{7}$ Department of Physics \& Astronomy, University of California, Riverside, 900 University Ave., Riverside, CA 92521, USA}

\begin{abstract}
We identify and analyze the little red dots (LRDs) in the AMBRA cosmological hydrodynamic simulation, by producing mock observations of the galaxies and active galactic nuclei (AGN) between redshifts 5 and 8. We produce these mock observations with both a standard AGN emission model, and a ``gas-enshrouded'' model, and find that the presence of a gas-enshrouded AGN is instrumental to the reproduction of dim LRDs (F444W magnitude $>$ 26.0). 
We find a steeper decrease in LRD density between $z=5$ and $z=8$ within AMBRA 
than seen in observations, resulting in a relative underproduction of LRDs beyond $z\sim6$ in AMBRA. With the addition of unresolved AGN variability, the decline with redshift flattens, and the LRD redshift evolution in AMBRA becomes broadly consistent with other theoretical datasets, and closer to that seen in observations. We find that the LRDs in AMBRA are pre-existing black hole--galaxy systems that are undergoing an LRD phase, selected primarily by the brightness of the AGN relative to its host. While the exact duration of these LRD phases is not possible to determine due to the unresolved nature of the AGN environment, we use the available time-series data to place limits on the lifetimes of the LRDs in AMBRA. We find that the vast majority of LRDs in AMBRA have lifetimes between $\sim3$ and $\sim 300$ Myr with LRD lifetime increasing with both black hole and host galaxy mass. The lower mass LRDs are more dependent on housing a gas-enshrouded AGN, and have shorter lifetimes ($\rm \sim30~Myr$). The higher mass LRDs are less sensitive to their AGN environment, but require more compact host galaxies, and have longer lifetimes ($\rm \sim100~Myr$).
\end{abstract}


\maketitle



\section{Introduction}
\label{sec:Intro}

One of the many discoveries from early JWST observations was the population of previously unknown objects known as ``little red dots'' \citep[LRDs;][]{Matthee_2024, Labbe_2023, Kocevski_2023, Harikane_2023, Furtak_2023, Kocevski_2025}. These objects were originally characterized by compact morphology, red rest-frame optical color, flat to blue rest-frame UV color, and broad emission lines. Early analysis of these objects attempted to determine their nature, producing two classes of models --- compact star-forming galaxies, and reddened AGN \citep{Baggen_2023, Barro_2023, Labbe_2023, Maiolino_2024_JADES, Guia_2024}.

Since their discovery, a significant amount of research has been dedicated to investigating the properties of these LRDs, ranging from follow-up observations to the production of new AGN emission models. The follow-up observations included JWST spectroscopy \citep{Kocevski_2023, Greene_2024, Matthee_2024}, X-ray \citep{Ananna_2024, Yue_2024, Maiolino_2025}, and infrared observations \citep{Casey_2025}. These observations identified several common features in the spectra of LRDs. Nearly all LRDs are not detected in X-rays \citep[][but see e.g. \citealt{Hviding2026, Kocevski_2025} for the few X-ray detected LRDs]{Akins_2025, Ananna_2024, Yue_2024, Maiolino_2025, Pacucci_Narayan_2024}, have significant broad line Balmer emission \citep{Wang_2024, Ma_2024}, feature a large Balmer break \citep{Setton_2024, Labbe_2024}, and have very minimal infrared \citep{Casey_2025} and radio \citep{Perger_2025} emission. 
Photometric and spectroscopic monitoring has additionally found the population to be, at most, weakly variable (\citealt{KokuboHarikane2024, Zhang2025, Furtak2025}; see also \citealt{Secunda_2026}), which is difficult to reconcile with an ordinary accretion-disk-dominated AGN. 

Much of the recent progress has come from 
exceptionally deep observations of individual objects. Strongly lensed objects such as the LRD Abell2744-QSO1 at $z = 7.04$ \citep{Furtak_2023, Furtak_2024} are magnified enough to allow the calculation of robust size constraints, the detection of a non-stellar Balmer break \citep{Ji2025}, 
and a direct dynamical black hole mass measurement of $M_{\rm BH} \sim 5\times10^{7}\,M_{\odot}$ 
\citep{Juodzbalis2026}. Comparably detailed analyses of MoM-BH*-1 \citep{Naidu_2025}, The Cliff \citep{deGraff_2025}, and CAPERS-LRD-z9 \citep{Taylor_2025} have reached similar conclusions. In parallel, searches at lower redshifts have found objects that share some of the defining LRD properties: V-shaped continua, broad Balmer lines, and in some cases Balmer absorption \citep{Lin2025, Ji2026, Ding2026, Rinaldi2025, Hviding2026}. While the exact relationship between these low-redshift analogs and the high-redshift LRDs is still uncertain, we can obtain much higher fidelity observations of them, which enables analysis that is impossible for sources at $z \geq 5$. 

While our understanding of the observational features of LRDs has continued to deepen, the combination of these features has remained difficult to reconcile with existing models of galaxy formation, stellar populations, and AGN emission. Dust-attenuated sources, both stellar and AGN, would likely be far brighter in the IR than observed LRDs due to dust reprocessing \citep{Casey_2024, Casey_2025}. Traditional AGN emission models without significant dust attenuation would not produce the visible red colors of observed LRDs, and may produce more X-ray emission than is allowed by current detection limits. Dense stellar sources would generally lack the broad emission lines seen in LRD spectra, but may be able to produce them under specific circumstances \citep{Perez_Gonzalez_2024, Baggen_2024, Guia_2024, Hviding_2025}. 

Some recent results have placed an increasing amount of importance on the Balmer break as the characteristic feature of the LRD spectrum \citep{Liu_2025}, and a variety of works have proposed AGN-centric models with spectral energy distributions (SEDs) that differ significantly from those based on catalogs of low-redshift quasars \citep{Shen_2020}. Two broad approaches have been taken. The first derives an adjustment to the AGN bolometric corrections empirically, extrapolating a possible AGN SED from the observed emission and detection limits of the brightest LRDs \citep{Greene_2025}. The second uses photoionization modeling to produce AGN spectra and fit the observed LRD spectra directly \citep{Inayoshi_2025, Naidu_2025, deGraff_2025, Taylor_2025, Pacucci_2026_DCBH}. These fits require atypical properties for the AGN and its surrounding gas: the AGN must be enshrouded within a very dense, dust-poor cloud of gas whose Balmer break, rather than dust, produces the red rest-optical color, while the AGN itself supplies the broad emission lines. This configuration simultaneously suppresses escaping X-ray emission and, by removing the need for heavy dust attenuation, keeps the predicted dust re-emission below the observed infrared upper limits. These models are also able to match both the continuum shape and the Balmer line profiles of LRDs simultaneously \citep{Ji2025, Taylor_2025, Rusakov_2025, deGraaff_2025_BHstar, Sun2026}. The origin of these dense gas environments is still uncertain. Some models like those in \citet{Inayoshi_2025} assume a short-lived influx of dense gas into the AGN environment. Others posit that they arise from recently formed direct collapse black holes (DCBHs) that are still ensconced in the gas cloud they formed within \citep{Naidu_2025, Pacucci_2026_DCBH}. 

These recent developments suggest that galaxies hosting bright AGN embedded in dense gas may represent much of the observed LRD population \citep{Sun2026}. In this case, there are still a number of very important questions about the nature of these objects that are unresolved. Are the black holes significantly overmassive relative to their hosts \citep{Pacucci_2023, Jones_2025}, and are they accreting below, near, or well above the Eddington rate? What is the origin and duration of the gas-enshrouded phase? If most LRDs have host galaxies, what are their properties, and how important are they to the source appearing as an LRD? Are all LRDs actually a single class of objects, or are they a heterogeneous population with similar observable characteristics? What drives the redshift evolution of the LRD population, which rises from $z \sim 9$ to $z \sim 5$ then declines? 

Additionally, the study of LRDs is closely intertwined with investigations of the population of potentially ``overmassive'' black holes observed at similarly early epochs, some of which have been identified as LRDs \citep{Pacucci_2023, Durodola_2024, Taylor_2025b, Furtak_2024, Jones_2025, Gupta_2026}. While there is the possibility that the masses of these objects have been overestimated \citep{Greene_2025, Brooks_2024, Brooks_2025, Rusakov_2025, Sun_2026, Curtis_2026}, it remains unclear which interpretation is correct. Further exploration of the physical conditions capable of producing such overmassive black holes, and whether these conditions also give rise to LRD populations consistent with observations, may help resolve this ambiguity.

An additional piece of the puzzle regarding the nature of LRDs was recently uncovered, with the discovery of abundant populations of nearby blue companion galaxies, at distances $\lesssim 5$ kpc from the central LRD \citep{Baggen_2026, Barger_2026}. These blue companions are detected around bright LRDs at rates that are overwhelmingly higher than in standard, control galaxies, matched in redshift, luminosity, and compactness \citep{Pacucci_Urry_2026}. This finding strongly suggests that the phenomenon of the LRDs is physically linked to the presence of these nearby blue companions.

Cosmological simulations can help answer many of these questions. They provide very large datasets of self-consistent black holes and host galaxies that include relevant astrophysical processes such as gas cooling, star formation, and stellar feedback, along with black hole seeding, accretion, and feedback. These objects serve as excellent test beds for proposed LRD models, provide a statistically significant sample for population-level analysis, directly present important black hole and galaxy properties that must be inferred for observed objects, and allow for the analysis of individual objects as they evolve through time. 

\begin{figure*}
    \centering
    \includegraphics[width=2.0\columnwidth]{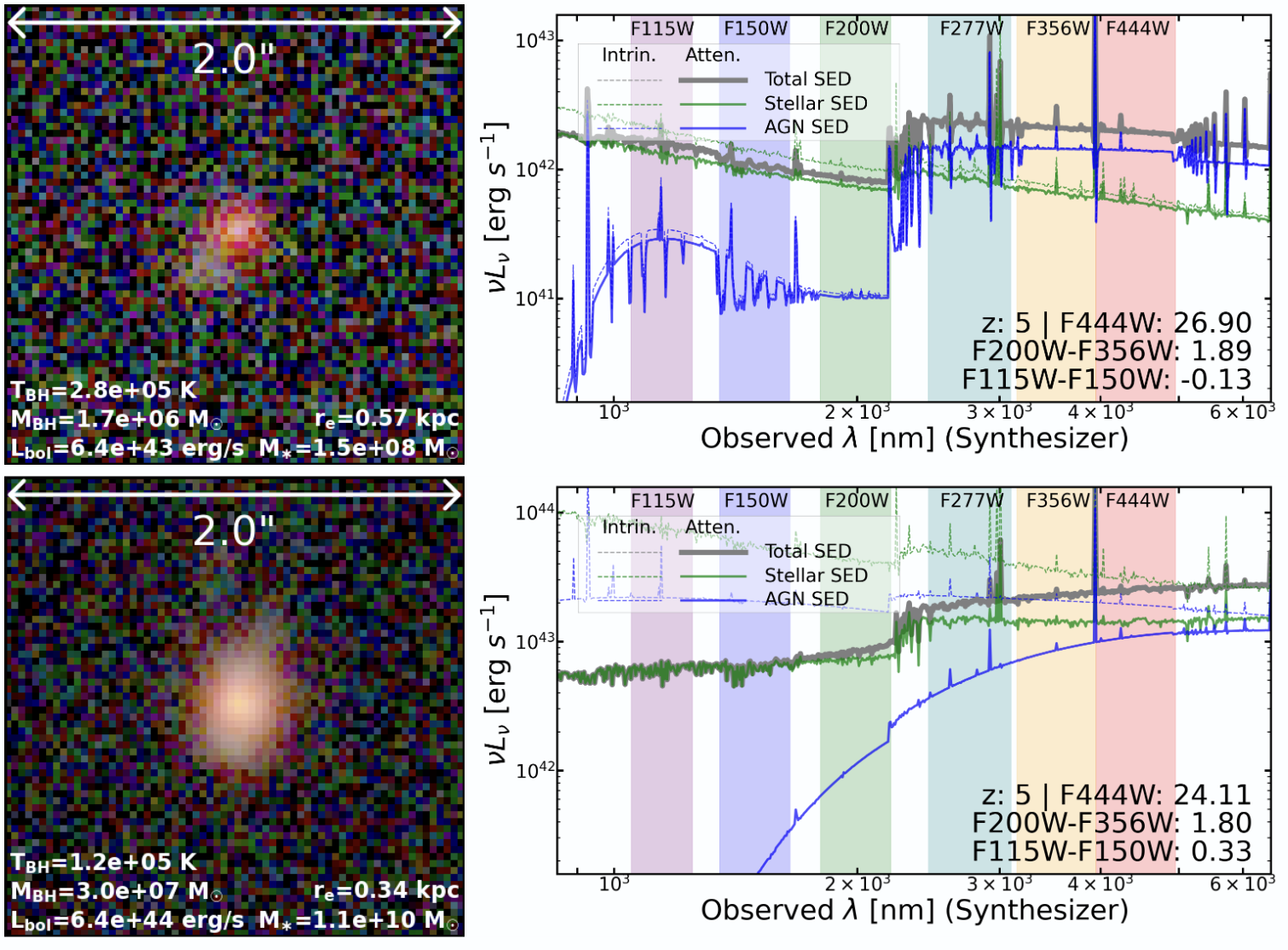}
    \caption{Mock JWST images and spectra of two of the little red dots found in the \ambra\ simulation. The top row shows a source that is an LRD with the gas-enshrouded AGN emission model, and the bottom row shows a source that appears as an LRD with the standard AGN emission model. The images include text with the physical properties of the source, including the mass, bolometric luminosity and characteristic temperature of the black hole, and the stellar mass and half-mass radius ($r_e$) of the host galaxy. The spectra include the observed F444W magnitude, and colors used for LRD identification.}
    \label{fig:Ambra_Mock_LRD_Obs}
\end{figure*}

The chief limitation of this approach is that cosmological simulations do not model all of the astrophysical processes required to predict what an LRD would look like. The production of emission from AGN and stars, and its attenuation by dust, must be added in post-processing. For the AGN, this is primarily done via radiative transfer modeling with software such as \texttt{cloudy} \citep{cloudy}, coupled to the black hole properties taken from the simulation. Because that radiative transfer occurs on length scales far below the resolution of the simulation, the aspects of the model that matter most for the gas-enshrouded scenario --- the volume density, column density, and turbulence of the gas cloud immediately surrounding the AGN --- cannot be inferred from the simulation and must instead be imposed as free parameters. Similarly, stellar emission is obtained by coupling the properties of each star particle to a stellar population synthesis model, and attenuating that emission based on the intervening gas particles. 

In a previous work, we produced mock observations of the galaxies and AGN in the \astrid\ simulation \citep{astrid_BHs, astrid_galaxy_formation, Ni_2025, Zhou_2025}, identified a limited number of LRDs, and analyzed their properties \citep{LaChance_2025}.

We then compared those results to the LRD population in a realization of the \brahma\ simulation suite~\citep{Bhowmick_2024, Bhowmick_2024c}, in which a gas-based seeding prescription combined with a black hole repositioning scheme leads to rapid, merger-driven mass assembly and a large population of black holes that are overmassive relative to their host galaxies. Despite hosting the most overmassive black holes of the simulations we have considered, \brahma\ only approached the observed LRD abundances under the assumption of a very high duty cycle for the gas-enshrouded AGN phase \citep{BRAHMA_LRDs}. The comparison between \astrid\ and \brahma\ was limited by the smaller volume of \brahma\ and by the many modeling differences between the two simulations, which left the role of the seeding model entangled with that of the other astrophysical models.

In this work, we shift our focus to the newly produced \ambra\ \citep{AMBRA} simulation. This simulation is nearly identical to \astrid\ (maintaining the volume, mass resolution, initial conditions, and all other astrophysical models), with the primary difference being the use of a gas-based black hole seeding prescription similar to those used in the \brahma\ simulation suite. This addresses many of the limitations of our analysis of the \brahma\ simulation by providing a significantly larger volume ($\sim 1000\times$ larger), and allowing all differences from the \astrid\ simulation to be attributed to the black hole seeding model. Notably, \ambra\ retains the subgrid dynamical friction model of \astrid\ for black hole dynamics, rather than the repositioning scheme of \brahma. As a result, \ambra\ still produces overmassive black holes, but produces fewer of them, and with lower masses than those in \brahma\ (see sections \ref{subsec:LRD_identification} and \ref{subsec:BH_gal_props}). 
In addition, we perform a time-domain analysis of the LRDs present in \ambra, tracking individual objects from $z=10$ to $z=5$ to monitor their properties and LRD lifespans.

This paper is organized as follows: A description of the simulations used in this work, the methods for creating mock observations, and identifying little red dots are presented in section \ref{sec:methods}. We analyze the properties of the little red dots within \ambra\ and their evolution across time in section \ref{sec:results}. We include a discussion of our findings, including the relevant assumptions, and potential for future work in section \ref{sec:discussion}. We conclude the paper with a summary of our results and the implications for the open questions regarding little red dots in section \ref{sec:Summary}.

\begin{figure*}
    \centering
    \includegraphics[width=2.0\columnwidth]{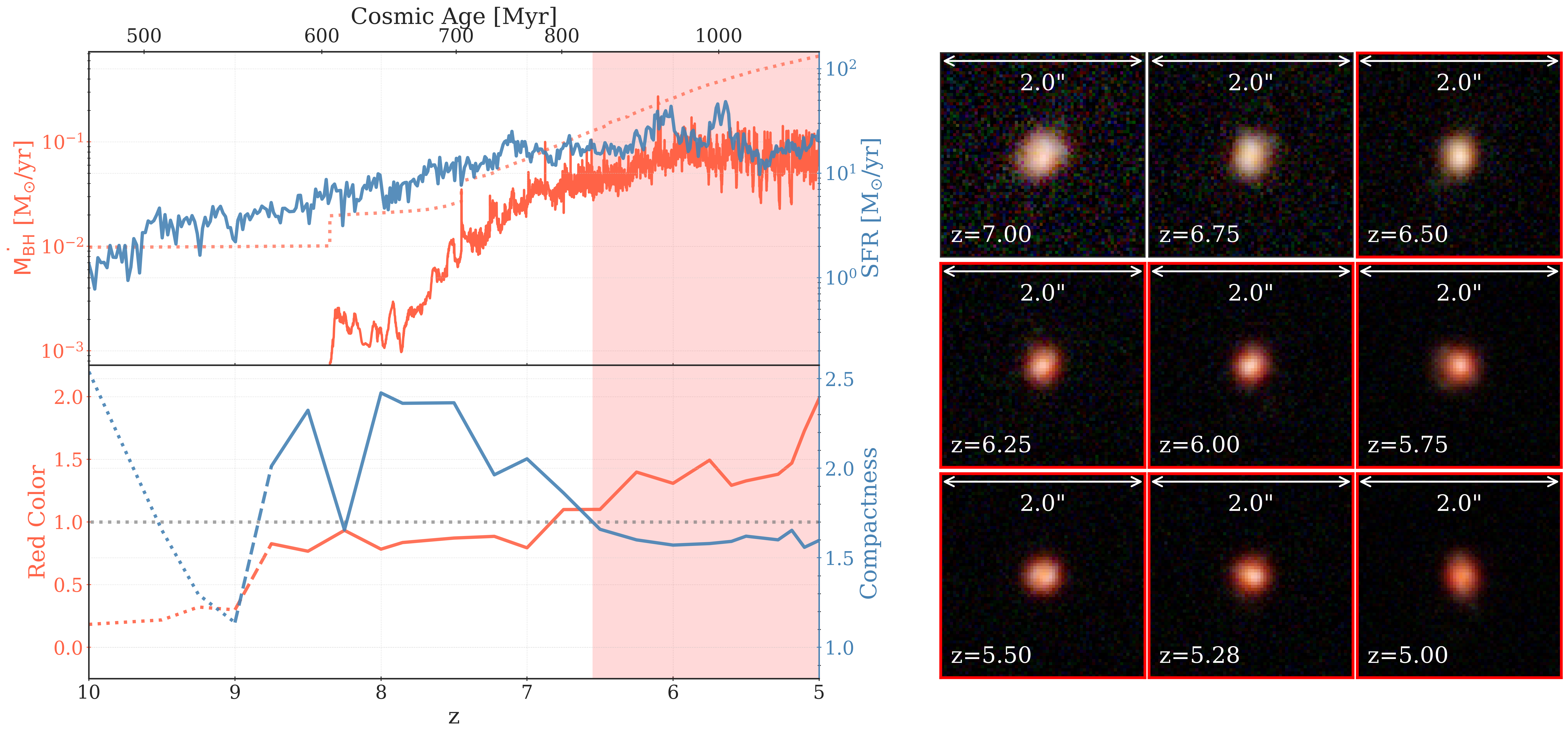}
    \includegraphics[width=2.0\columnwidth]{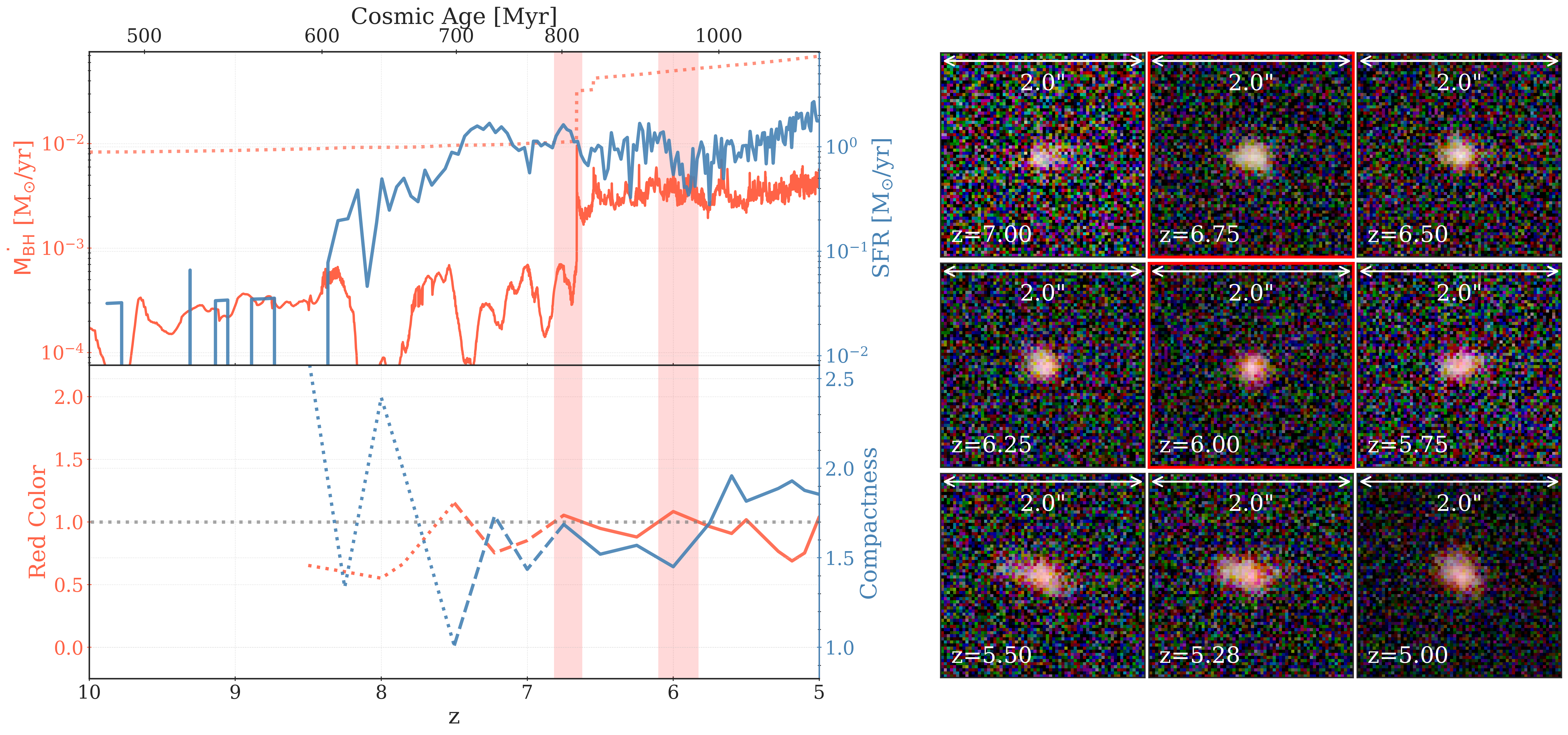}
    \caption{Time-series data for two of the little red dots found in the \ambra\ simulation, one per row. For each source, the left-hand side shows the history of multiple physical and observable properties, and the right-hand side shows a grid of time-series mock images. Within each history plot, the upper panel shows the physical properties of the source --- specifically the black hole accretion rate history in orange, and the star formation rate history in blue --- and the lower panel shows the observable properties relevant to LRD identification, with the \redcolorLabel\ in orange, and the compactness in blue. The horizontal dotted line in the lower panel marks the \redcolorLabel\ threshold of 1.0, and the compactness threshold of 1.7. We shade the periods where the source is identified as an LRD in red. The time-series images include observations before, during, and after (if applicable) a period where the source would appear as an LRD, and observations where the source meets the LRD criteria are outlined in red. All of the images include the background and Poisson noise described in section \ref{subsec:segmgentation}; the noise is far less apparent in the images of the top source simply because it is intrinsically much brighter than the bottom source. These are the same two sources whose evolution is shown as the green (top source) and gold (bottom source) tracks in figures \ref{fig:MBH_Mstar}--\ref{fig:sSFR_r_half}.}
    \label{fig:histories_w_images}
\end{figure*}

\section{Methods}
\label{sec:methods}

\subsection{The \ambra\ Simulation}
\label{subsec:ambra}

The \ambra\ simulation (\textbf{A}STRID with \textbf{M}BH seeding from \textbf{BRA}HMA) \citep{AMBRA} is a cosmological hydrodynamical simulation designed to study the early assembly of massive black holes. It combines the large volume and statistical power of the \astrid\ simulation \citep{astrid_BHs, astrid_galaxy_formation, Ni_2025, Zhou_2025} with a gas-based black hole seeding prescription taken from the \brahma\ simulation suite \citep{Bhowmick_2024, Bhowmick_2024c}. Aside from this seeding model, \ambra\ is identical to \astrid\ in every way, sharing its volume, mass resolution, initial conditions, and all of its astrophysical and feedback models. Any differences between the black hole populations of the two simulations can therefore be attributed directly to the seeding model.

\ambra\ inherits its numerical framework from \astrid. It is run with the Smoothed Particle Hydrodynamics (SPH) code \texttt{MP-GADGET} \citep{Feng2018_MPGadget}, using a standard set of cosmological parameters based on the Planck survey \citep{Planck_2020}. The simulation volume is $[250\,h^{-1}\,{\rm Mpc}]^3$, with an initial particle load of $\rm 2 \times 5500^{3}$ particles. The resulting mass resolution is similar to the Illustris, TNG100 and EAGLE simulations, with a simulation volume larger than Illustris TNG300 (IllustrisTNG: \citealt{Springel_2018, Nelson_2018, Marinacci_2018, Naiman_2018}; EAGLE: \citealt{Schaye_2015}). Star formation is performed with the model described in \citet{Feng_2016}, which is based on the model originally developed in \citet{Springel_2003}, with an additional correction to handle the formation of molecular hydrogen, and corresponding adjustments to the star formation model in low metallicity environments per \citet{Krumholz_2011}. Gas cooling follows the prescription of \citet{Katz_1996}, with an implementation of dense gas self-shielding per \citet{Rahmati_2013}. The full description of the stellar and gas processes is detailed in \citet{astrid_galaxy_formation} and \citet{astrid_BHs}. \ambra\ has been evolved to $z=5$, which sets the low redshift limit of our analysis.

The black hole modeling beyond seeding is likewise unchanged from \astrid. Black holes accrete via a modified Bondi--Hoyle prescription with a boost factor of 100 and a radiative efficiency of $\eta = 0.1$, and mildly super-Eddington accretion is permitted, up to twice the Eddington limit. The AGN feedback model includes both thermal and kinetic feedback modes. Black hole dynamics are captured with a subgrid dynamical friction model \citep{Chen_2022}, with black hole mergers occurring only between close, gravitationally bound pairs. 

The black hole seeding model is where \ambra\ departs from \astrid. In \astrid, halos are seeded with initial black holes of mass $\sim4.4\times10^{4}$--$4.4\times10^{5}\,M_{\odot}$, drawn from a power-law distribution, only once both their total halo mass and stellar mass exceed thresholds of $7.4\times10^{9}\,M_{\odot}$ and $3\times10^{6}\,M_{\odot}$, respectively. \ambra\ removes these mass thresholds entirely, replacing them with a seeding criterion based on the gas properties of early halos, following the \brahma\ simulation suite. In the \brahma\ framework, black hole seeds are placed in halos that contain a critical mass of dense, star-forming, metal-poor ($Z < 10^{-4}\,Z_{\odot}$) gas, optionally subject to additional criteria such as a minimum Lyman--Werner (LW) flux. Among the variants explored in the \brahma\ suite, only those requiring the smallest mass of dense, metal-poor gas ($\rm\sim 5 M_{seed}$) and little to no LW flux are able to reproduce massive JWST-detected black holes such as GN-z11 \citep{bhowmick_2026}. \ambra\ adopts a seeding model in this regime: seeds are placed in any resolved halo (an implicit halo mass floor of $\sim3\times10^{8}\,M_{\odot}$) that contains a minimum mass of star-forming, metal-poor gas, which is set to a single gas particle ($\sim1.9\times10^{6}\,M_{\odot}$). The seeding prescription used in \ambra\ does not include any criterion on the LW flux. As such, this seeding model is most aligned with seed formation via runaway collisions in nuclear star clusters \citep{Davies_2011, Lupi_2014, Kroupa_2020, Das_2021b, Das_2021a, Kritos_2023} and population III stellar remnants that undergo rapid hyper-Eddington accretion \citep{Fryer_2001, Madau_2001, Xu_2013, Smith_2018, Mehta_2026} rather than DCBH scenarios which would rely on LW radiation. The seed masses are drawn from the same power-law distribution used in \astrid. 

This change to the seeding model produces a substantially different black hole population than was found in \astrid. In \ambra, the first seed forms at $z \approx 26$, compared to $z \approx 17$ in \astrid, and by $z=8$ the number density of black holes with masses of $10^{5}$--$10^{7}\,M_{\odot}$ is more than an order of magnitude higher than in \astrid. The abundant early seeds also allow for merger-driven growth, with mergers accounting for $\sim50\%$ of the mass of the most massive black holes at $z\gtrsim11$, enabling \ambra\ to reproduce massive JWST-detected black holes, such as those in GN-z11 and CEERS-1019, that \astrid\ cannot \citep{AMBRA}. For context, the original \astrid\ seeding model produces a black hole population at $z \sim 3$--$7$ that is in agreement with pre-JWST predictions from X-ray observations \citep{Willott_2010, Ueda_2014}, and in our previous mock observation analysis of \astrid\ we identified only a small population of LRDs \citep{LaChance_2025}. In contrast, the \brahma\ realizations run with lenient gas-based seeding host black holes that are significantly overmassive relative to their host galaxies, and in \citet{BRAHMA_LRDs} we found that this significantly increased the LRD number density. That analysis was limited, however, by the small $[36\,{\rm Mpc}]^3$ volume of the \brahma\ box, and by the many modeling differences between \brahma\ and \astrid\ beyond seeding. \ambra\ addresses both limitations. As both \astrid\ and \brahma\ are important points of reference for the \ambra\ simulation, we include our previous results from both \astrid\ \citep{LaChance_2025} and \brahma\ \citep{BRAHMA_LRDs} as points of comparison throughout this work.

\subsection{Mock Observation Pipeline}
\label{subsec:Obs_pipeline}

For this work we utilize a mock observation pipeline that is similar to the methods we employed in \citet{BRAHMA_LRDs, LaChance_2025}, but with a few adjustments. The underlying emission modeling remains the same as \citet{BRAHMA_LRDs}, but we have re-implemented that pipeline within the mock observation software \texttt{SYNTHESIZER} \citep{Lovell_2025, Roper_2026}. Additionally, in order to facilitate the time-series analysis detailed in section \ref{subsec:histories}, we have added image segmentation into our pipeline. 
This pipeline produces mock observations like those shown in figure \ref{fig:Ambra_Mock_LRD_Obs}.

\subsubsection{Stellar emission and dust attenuation}
\label{subsec:stellar_emission}
We begin by calculating the emission from each star particle. Their intrinsic SEDs are assigned based on their properties, using the Binary Population and Spectral Synthesis model \citep[BPASS version 2.2.1;][]{Stanway2018}. Once calculated, we apply dust attenuation to these stellar spectra. We calculate the attenuation using a simple power law dust model: 
\begin{equation} \label{eq_tau}
\tau_{\rm ISM}(\lambda) = -\kappa_{\rm ISM} \Sigma(x,y,z) \left(\frac{\lambda}{0.55\,\mu{\rm m}}\right)^{\gamma}
\end{equation}
where $\tau_{\rm ISM}(\lambda)$ is the optical depth, $\kappa_{\rm ISM}$ is a tuning parameter, which we assign a value of $10^{4.1}$ based on the calibration done in  \citet{astrid_galaxy_formation}, $\Sigma(x,y,z)$ is the metal surface density of each star, and $\gamma$ is the slope of the dust model. We set $\gamma = -1.0$ to maintain consistency with the dust calibration in \citet{astrid_galaxy_formation}. This results in a dust curve which falls between the Small Magellanic Cloud model \citep{Pei_1992} and the `starburst' model \citep{Calzetti_2000}. In an update from prior works, we calculate $\Sigma(x,y,z)$ for each star by summing the contributions of all gas particles whose smoothing kernels overlap the line of sight. This is the same method we used to calculate the metal surface density of the AGN in \citet{BRAHMA_LRDs}.

In addition to the dust attenuation from the interstellar medium (ISM) described above, we also allow young stars (age $\leq 10$ Myr) to produce nebular emission, and experience birth cloud dust attenuation. We calculate our birth cloud attenuation using the procedure detailed in \citet{Vijayan_2021}, and the nebular emission with the process described in \citet{wilkins_2020}.

\begin{figure*}
    \centering
    \includegraphics[width=2.0\columnwidth]{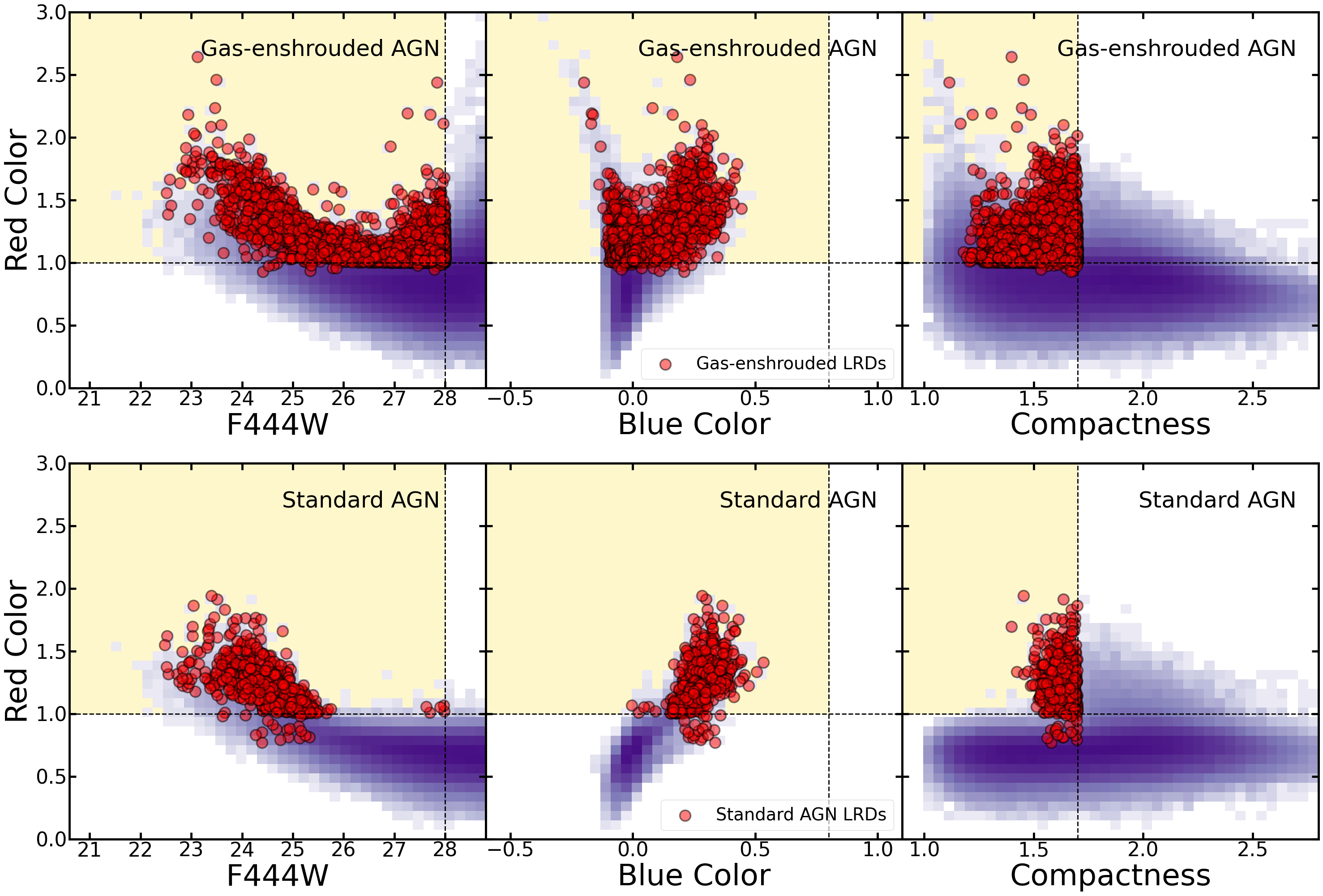}
    \caption{Color-Color and Color-Magnitude plots for the colors and magnitudes used in our little red dot criteria. The top row shows the properties of the \ambra\ sources when utilizing the gas-enshrouded AGN model, and the bottom row contains the results when a standard AGN model is adopted.
    The left panels show the sources' \redcolorLabel\ vs their observed F444W magnitude. \redcolorLabel\ is the F200W-F356W color for objects in the $z=5$, 6 snapshots, and the F277W-F444W color for objects in the $z=7$, 8 snapshots. These correspond to colors used in the Red1 and Red2 LRD criteria to determine if an object is red enough to be an LRD. The middle panels show \redcolorLabel\ vs \bluecolorLabel. \bluecolorLabel\ is the F115W-F150W color for objects in the $z=5$, 6 snapshots, and the F150W-F200W color for objects in the $z=7$, 8 snapshots. These correspond to the colors used in the Red1 and Red2 LRD criteria to confirm that an object has the spectral break of an LRD. The right panels show the \redcolorLabel\ vs the observed compactness of the sources. In order for a source to be considered an LRD it must be above the horizontal dashed line, and to the left of the vertical dashed line (falling within the highlighted regions) in all three panels. The 2D histogram shows the population of sources in \ambra. \icontext. All panels include all of the sources in the snapshots we analyzed ($z=5$--$8$).}
    \label{fig:LRD_criteria}
\end{figure*}

\subsubsection{AGN SED production}
\label{subsec:AGN_SED}

We utilize both a ``standard'' AGN and ``gas-enshrouded'' AGN emission model throughout this work, with a focus on the ``gas-enshrouded'' model. We produce both of these models in the exact same fashion as \citet{BRAHMA_LRDs}. They are created using the photoionization code \texttt{cloudy}. Both versions of the AGN emission model use the same underlying incident AGN model, which is produced using the \texttt{cloudy} AGN command, with default parameters. These are $\alpha_{\rm ox}=-1.4$ (the X-ray to UV ratio); $\alpha_{\rm UV}=-0.5$ (the slope of the low energy side of the Big Bump feature); and $\alpha_{\rm X}=-1.0$ (the X-ray slope). We produce AGN spectra for characteristic temperatures ranging from $10^{4.0}$~K to $10^{6.0}$~K. These characteristic temperatures depend on the mass and accretion rate of the black hole, and correspond to the peak of the ``Big Bump'' feature of the incident AGN spectrum.

Generally, the temperature of the AGN accretion disk is 
\begin{equation} \label{T_BH}
T_{\rm BH}(r) = \left(\frac{3 G M_{\rm BH}\dot{M}_{\rm BH}}{8\pi\sigma r^3}\right)^{1/4} (1-\sqrt{r_{\rm ISCO}/r})^{1/4}
\end{equation}

Here, $T_{\rm BH}(r)$ is the temperature of the accretion disk at distance $r$ from the black hole, $\dot{M}_{\rm BH}$ is the accretion rate of the black hole, and $r_{\rm ISCO}$ is the radius of the innermost stable circular orbit  ($r_{\rm ISCO} = \frac{6 G M_{\rm BH}}{c^2}$). The maximum temperature is found at $r = \frac{49}{36} r_{\rm ISCO}$, and we use the temperature at this radius as the characteristic temperature for the AGN SED. 

In addition to the parameters of the incident spectrum, we also choose values for the gas cloud that the incident spectrum is processed through. For both models we use a metallicity of $10\%$ of the solar value, and choose an inner radius of $5 \times 10^{16}$~cm. These templates are produced for an AGN with a bolometric luminosity of $10^{43.5}~\rm erg~s^{-1}$, and scaled to match the luminosity of the AGN from the simulation. Notably, scaling the template SED effectively also increases the inner cloud radius by the same factor, resulting in the ionization parameter $U$ being held constant for all AGN of a given temperature.

We calculate the luminosity of each black hole according to \begin{equation} \label{eq_L_bol}
L_{\rm bol} = \eta\dot{M}_{\rm BH} c^2
\end{equation}

where $\eta$ is the radiative efficiency of the accretion onto the black hole, which we assign a value of 0.1 for \ambra. 

We apply the same ISM dust attenuation to the AGN as we used for the stars, with one exception. We employ a dust law with a steeper slope of $\gamma = -2.0$ for the AGN, compared to $\gamma = -1.0$ for the stars. As we will discuss in sections \ref{subsec:LRD_identification} and \ref{subsec:BH_gal_props}, we find that dust attenuation of the AGN is a fairly minor effect, only relevant to the production of brighter LRDs, as the dimmer LRDs in our sample are reddened via another mechanism.

The standard and gas-enshrouded AGN spectra differ in a few important properties of the gas cloud used in \texttt{cloudy} --- namely, the hydrogen volume and surface density, in addition to the presence and strength of turbulent motion within the gas clouds. The standard AGN utilizes a hydrogen volume density of $\rm log(n(H) / cm^{-3})= 9.0$, and a column density of $\rm log(N(H) /cm^{-2}) = 23.0$. The gas-enshrouded AGN model increases both of these values to $\rm log(n(H) / cm^{-3})= 9.5$, and $\rm log(N(H) /cm^{-2}) = 24.5$, and introduces a turbulent velocity of $400$~km~s$^{-1}$ in order to reproduce the widths of the Balmer absorption features and the sloped Balmer breaks seen in observed LRD spectra \citep{Naidu_2025, Taylor_2025, Ji2025, deGraff_2025}. This gas-enshrouded model is similar to those used in previous works \citep{Inayoshi_2025, Naidu_2025, Taylor_2025, deGraff_2025, Jeon_2025}, with the specific values chosen as they fall near the middle of the explored ranges, and produce the largest number of LRDs in both \astrid\ and \brahma\ in \citet{BRAHMA_LRDs}.

We apply both the standard and gas-enshrouded AGN models to every black hole in the simulation, rather than attempting to assign one or the other to individual sources based on their properties. The physical conditions that determine whether an AGN is gas-enshrouded --- the density, column, and turbulence of the gas within the environment of the black hole --- are many orders of magnitude below the resolution of \ambra, and the resolved properties of the host galaxy and black hole (e.g. gas density, accretion rate, or Eddington ratio) are not known to be reliable predictors of whether such a phase is occurring. The origin and duration of the gas-enshrouded phase also remain uncertain, with proposed lifetimes ranging from $\lesssim 1$ Myr \citep{Inayoshi_2016, Takeo_2020, Shi_2023} to tens of Myr \citep{Coughlin_2024, Begelman_2026, Sun2026}, and estimated duty cycles as low as $\sim 1\%$ \citep{Sun2026}. Any attempt to select which sources are enshrouded would therefore be imposing an additional, poorly constrained model on top of the simulation. Instead, we treat the two models as bracketing cases. Applying the gas-enshrouded model to all AGN is equivalent to assuming a duty cycle of unity for the gas-enshrouded phase, and thus the resulting LRD population represents an upper limit on the number of LRDs that \ambra\ can produce via this channel. Applying the standard model to all AGN corresponds to a duty cycle of zero, and represents the LRD population that would be present without any contribution from the gas-enshrouded phase. The true LRD population in \ambra\ lies between these two limits, and its position depends on the duty cycle of the gas-enshrouded phase. Comparing the two also allows us to isolate which LRDs are dependent on the gas-enshrouded phase, and which would appear as LRDs regardless, a distinction we return to in sections \ref{subsec:LRD_identification} and \ref{subsec:LRD_lifecycle}.

\begin{figure*}
    \centering
    \includegraphics[width=2.0\columnwidth]{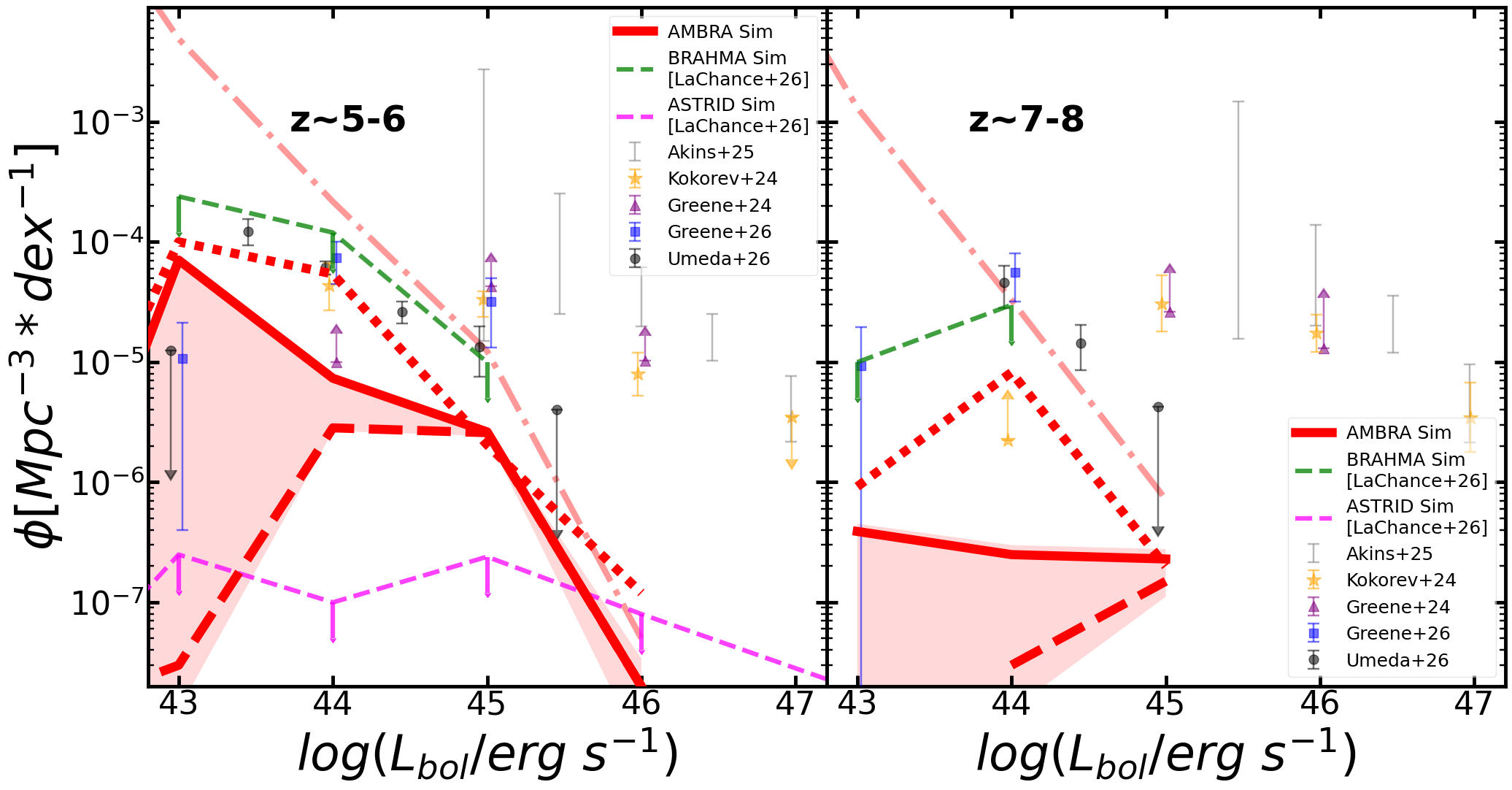}    
    \caption{The bolometric luminosity function of the black holes, and LRDs in the \ambra\ simulation. The left panel shows the results at redshifts $z=5$--$6$, and the right panel shows the results for redshifts $z=7$--$8$. The solid red line corresponds to the \ambra\ LRD population with gas-enshrouded AGN, the dashed red line corresponds to the \ambra\ LRD population with a standard AGN, and the dash-dotted line indicates the overall population of black holes in the simulation. The dotted red line is the \ambra\ LRD population with both the gas-enshrouded AGN model and the extra black hole luminosity stochasticity modeled per \citet{Lumina}, with $\sigma_{\rm bol} = \AGNstoch$ dex. \shadebetweentext. We also include the LRD populations we identified in our previous works on \astrid\ \citep{LaChance_2025} and \brahma\ \citep{BRAHMA_LRDs} as the magenta and green dashed lines respectively, and the observed LRD luminosity functions of \citet{Akins_2025, Kokorev_2024, Greene_2024, Greene_2025, Umeda_2026}.
    }
    \label{fig:BH_LF}
\end{figure*}

Finally, we consider the impact of AGN variability that is not captured by the simulation. The accretion rates recorded by \ambra\ are instantaneous Bondi--Hoyle estimates based on the gas properties resolved around each black hole, and therefore do not include the variability of AGN on the spatial and temporal scales below the resolution of the simulation. 
\citet{Lumina} account for this by adding an intrinsic stochasticity of black hole accretion at the sub-resolution level in the form of a log-normal scatter with a dispersion of $\sigma_{\rm bol}$. They find a dispersion of $\sigma_{\rm bol} = 0.3$ dex brings the AGN luminosity functions of LUMINA into agreement with observed quasar luminosity functions. Similar treatments, with scatters of $0.3$--$0.75$ dex, were used to reconcile the AGN luminosity functions of the EAGLE simulation \citep{Schaye_2015}, and the FLAMINGO simulation \citep{Schaye_2023} with observations \citep{Rosas_Guevara_2016, Ding_2026}. We present the impact of such a model on the LRD population of \ambra\ in figures \ref{fig:BH_LF}--\ref{fig:LRD_z_trend}, where we apply a log-normal scatter with $\sigma_{\rm bol} = \AGNstoch$ dex to the bolometric luminosity of each black hole before producing its AGN SED and mock observations. For reference, a \AGNstoch\ dex scatter is comparable to the amplitude of the resolved fluctuations in the accretion rate history of our example dim LRD (figure \ref{fig:histories_w_images}). Combined with the fact that this value falls near the middle of the range of dispersions used in other works, this gives us confidence that it is a reasonable choice for the unresolved variability. As this scatter is a statistical description of unresolved variability, rather than a physical prediction for any individual source, we perform our analysis of LRD properties, histories, and lifetimes on the dataset without the extra stochasticity included.

\subsubsection{Mock images and source extraction}
\label{subsec:segmgentation}

In order to produce the time-series mock observations described in section \ref{subsec:histories}, we must produce mock observations of galaxies in the intermediate snapshots, which do not have \texttt{subfind} catalogs. We begin this process with the $z=5$, 6, 7, and 8 snapshots, which do have full \texttt{subfind} catalogs. For each subhalo we produce a mock image which includes all of the particles associated with the parent Friends-of-Friends (FoF) halo, and is centered on the subhalo. We perform source identification and extraction on these images, and select the source which is closest to the expected position of the subhalo. When calculating the spectrum, photometry, and morphology measures associated with the galaxy, we only include the pixels that are identified as part of this source. For the intermediate snapshots, we create mock observations for each of the galaxies that are identified as LRDs based on the criteria outlined in section \ref{subsec:LRD_identification}. We center these mock images on the central black hole of each galaxy.

We create the mock images for JWST NIRCam filters F444W, F356W, F277W, F200W, F150W, F115W, and F090W. We produce these images with a drizzled pixel scale of $0.030''$ for all filters (original long and short wavelength filters have pixel scales of $0.063''$ and $0.031''$ respectively). We also apply a Gaussian point spread function (PSF) to the mock observation of each filter, with widths based on the empirical PSFs for each filter (e.g. FWHM $=0.145''$ for the F444W filter, and $0.040''$ for the F115W filter). In order to simulate the effects of background and Poisson noise, we add a uniform background with a flux based on the empirical average sky radiance observed in each filter. We convert the expected flux in each pixel into a number of detections based on pixel sensitivity, and an observation duration. We use an observation duration of 10,000 seconds in order to emulate the JWST Advanced Deep Extragalactic Survey \citep[JADES;][]{Eisenstein_2023}. We then apply Poisson noise to each pixel based on the number of expected detections, and convert those noisy values back to observed flux in each filter. This produces the background noise that is necessary to perform proper source detection, and image segmentation. 

We perform our source detection, and segmentation using \texttt{photutils} \citep{photutils}. We use a 1.5-sigma detection threshold for source detection. After initial source detection, we also perform source deblending, requiring any source to include at least 10 connected pixels. This allows us to properly separate our intended sources from any nearby galaxies, and removes any sources which would be difficult to isolate from their environment.

\subsection{Time-series mock observation and black hole histories}
\label{subsec:histories}

Using the above mock observation pipeline, we are able to produce mock observations of galaxies in the \ambra\ simulation roughly every $\Delta z = 0.25$. This allows us to produce a history of observable properties of the galaxies, including their magnitudes, colors, and sizes, and their underlying properties such as stellar mass, black hole mass, and star formation rate. Additionally, we utilize the detailed black hole information recorded by the simulation to produce black hole accretion histories for each of the LRDs present in \ambra. This allows us to track the lifecycle of each source which experiences an LRD window during this epoch in \ambra. We present two examples of these histories as tracks on several of the plots in this work (figures \ref{fig:MBH_Mstar}--\ref{fig:sSFR_r_half}), and examine the length of these LRD windows, and what might cause them to begin and end in section \ref{subsec:LRD_lifecycle}. We show an example of these histories in figure \ref{fig:histories_w_images}, which includes the observable properties that play a key role in determining if the source is an LRD, the underlying star formation, and black hole accretion histories that impact these observable properties, and mock images of the source before, during, and after it experiences its LRD window.

\begin{figure}
    \centering
    \includegraphics[width=1.0\columnwidth]{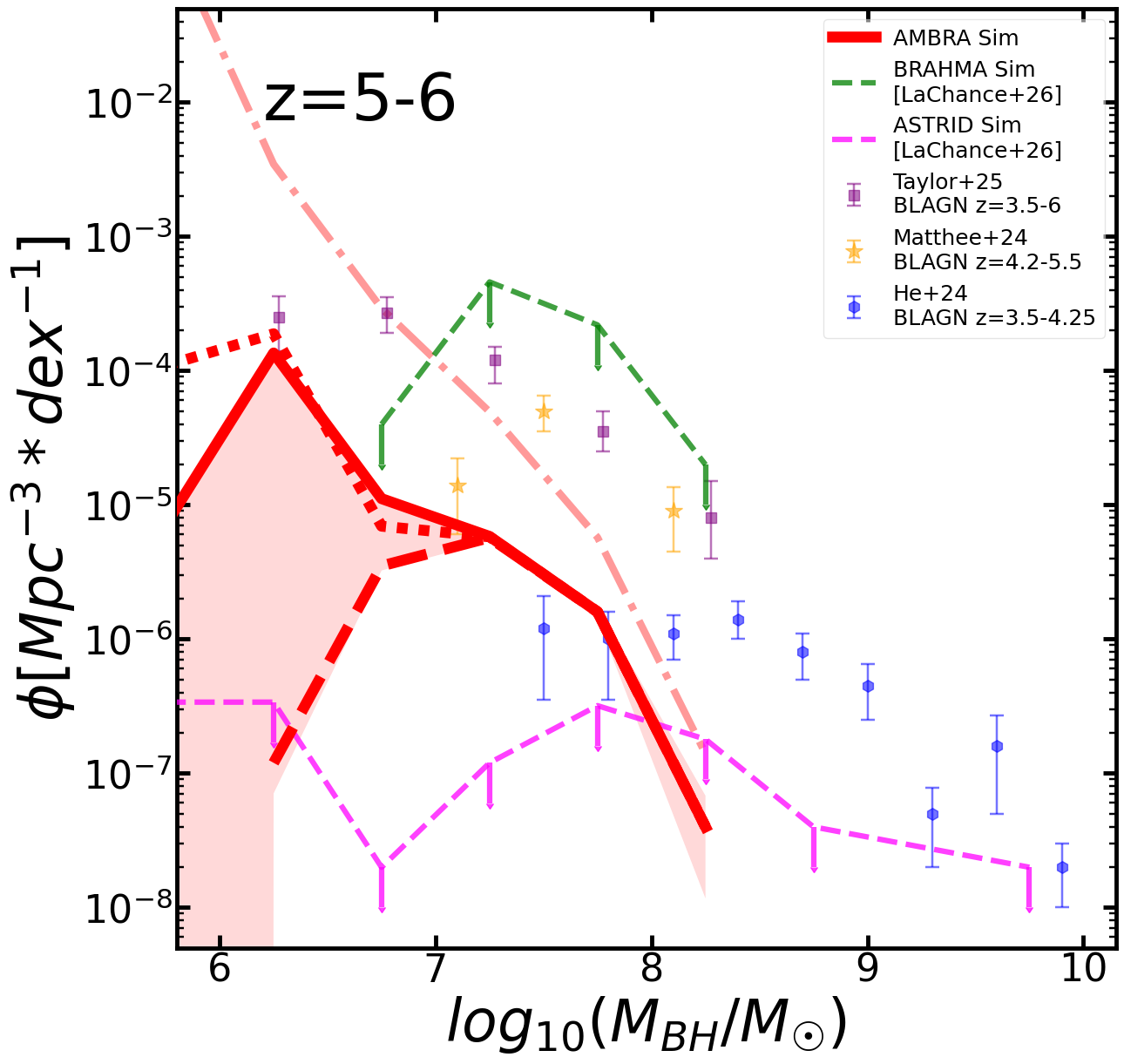}    
    \caption{The black hole mass function of the whole population, and the LRDs in the \ambra\ simulation. Here we only include the results for the $z=5$ and 6 snapshots to more closely align with the available observational results. The solid red line corresponds to the \ambra\ LRD population with gas-enshrouded AGN, the dashed red line corresponds to the \ambra\ LRD population with a standard AGN, and the dash-dotted line indicates the overall population of black holes in the simulation. The dotted red line is the \ambra\ LRD population with both the gas-enshrouded AGN model and the extra black hole luminosity stochasticity modeled per \citet{Lumina}, with $\sigma_{\rm bol} = \AGNstoch$ dex. \shadebetweentext. We also include the LRD populations we identified in our previous works on \astrid\ \citep{LaChance_2025} and \brahma\ \citep{BRAHMA_LRDs} as the magenta and green dashed lines respectively, and the observed broad-line AGN (BLAGN) mass functions of \citet{Taylor_2025b, Matthee_2024, He_2024}.
    }
    \label{fig:BH_MF}
\end{figure}

\section{Results}
\label{sec:results}

We apply the mock observation pipeline described in section \ref{subsec:Obs_pipeline} to the \ambra\ simulation, producing mock images and spectra for the underlying galaxy population in \ambra, allowing us to identify the LRDs that are present. We apply the full time-series data generation pipeline from section \ref{subsec:histories} to these LRDs in order to investigate how long they may appear as LRDs, and what physical processes may be responsible for the onset, and conclusion of these LRD periods.

\begin{figure*}
    \centering
    \includegraphics[width=2.0\columnwidth]{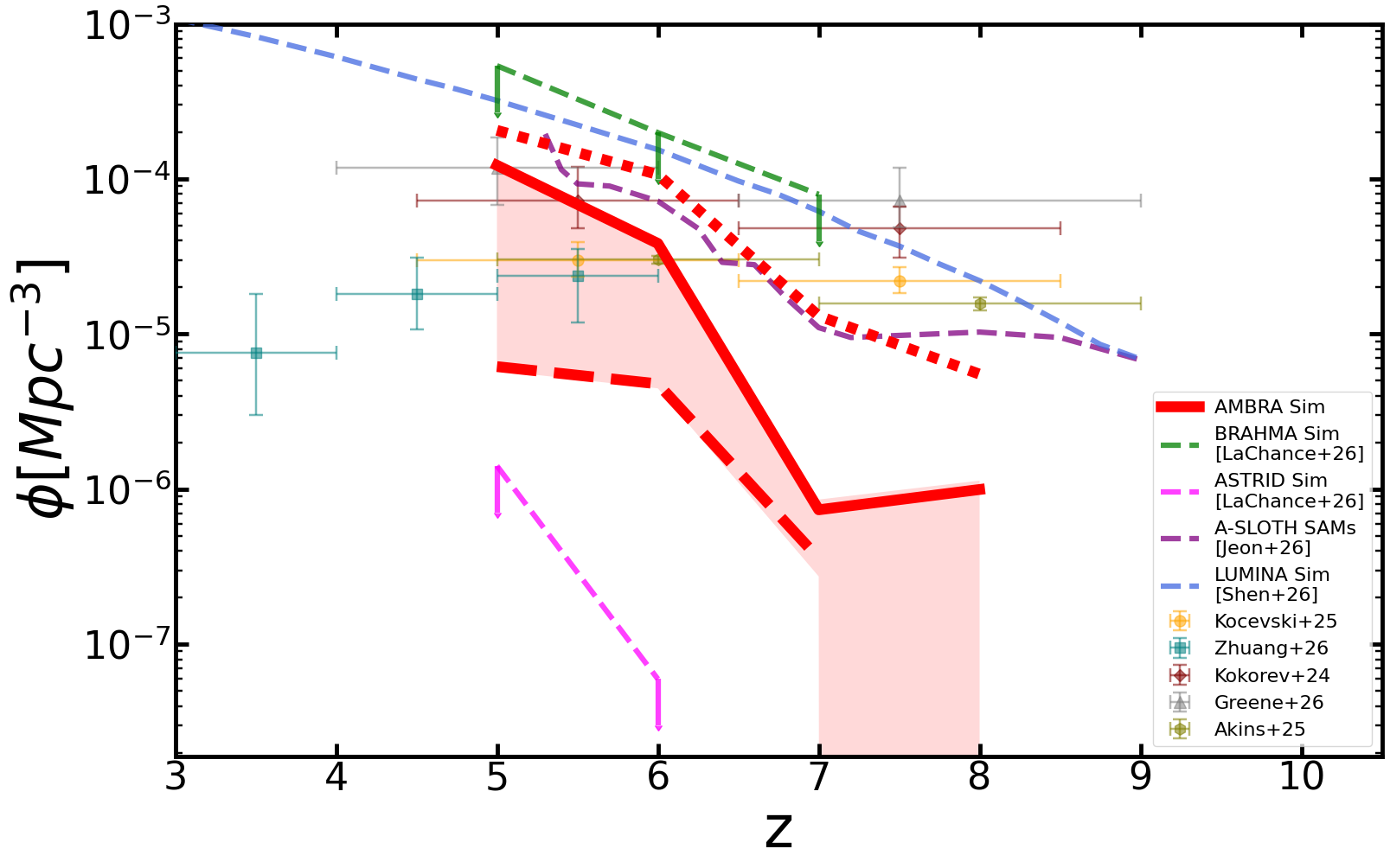} 
    \caption{The redshift evolution of the LRD population in \ambra, the LRD populations in our previous works on \astrid\ \citep{LaChance_2025} and \brahma\ \citep{BRAHMA_LRDs}, the LRD population in various observations \citep{Kocevski_2025, Zhuang_2026, Kokorev_2024, Greene_2025, Akins_2025}, and theoretical works (the heavy seed A-SLOTH SAMs of \citealt{Jeon_2025}, and the LUMINA simulation of \citealt{Lumina}). These other theoretical works select LRDs via their black hole properties rather than via mock observables (see section \ref{subsec:comparison_w_other_works}). The solid red line corresponds to the \ambra\ LRD population with gas-enshrouded AGN, and the dashed red line corresponds to the \ambra\ LRD population with a standard AGN. The dotted red line is the \ambra\ LRD population with both the gas-enshrouded AGN model and the extra black hole luminosity stochasticity modeled per \citet{Lumina}, with $\sigma_{\rm bol} = \AGNstoch$ dex. \shadebetweentext.
    }
    \label{fig:LRD_z_trend}
\end{figure*}

\subsection{LRD Identification and population}
\label{subsec:LRD_identification}

We use LRD selection criteria similar to those used by UNCOVER \citep{Greene_2024, Labbe_2025}, and to the criteria we used previously in \citet{BRAHMA_LRDs}. We apply a magnitude cut on the F444W emission of $m_{\rm F444W} < 28.0$ to ensure the objects are cleanly detected. Sources are then selected based on their colors, which must fit at least one of the Red1 or Red2 criteria,

\begin{align*}
\text{Red1 = (F115W-F150W < 0.8) and} \\
\text{(F200W-F356W > 1.0),} \\
\text{Red2 = (F150W-F200W < 0.8) and} \\
\text{(F277W-F444W > 1.0)}
\end{align*}

and their compactness in the F444W band, which must meet the following criterion:

$$\rm compact = f_{F444W}(0.4'')/f_{F444W}(0.2'') \leq 1.7$$

where $\rm f_{F444W}(0.4'')$ is the flux detected in the F444W band inside an aperture with a diameter of $0.4''$. Notably, we have removed the second redness criterion for both the Red1 and Red2 criteria ($\rm F200W -  F277W > 0.7$ and $\rm F277W -  F356W > 0.7$ respectively) in this work. The original intent of this criterion was to remove any sources which may have been strongly impacted by bright emission lines, rather than a red continuum. Our dataset does not contain these contaminating sources, which makes the criterion unnecessary for our sample. Additionally, it can introduce inconsistencies in the types of objects which are classified as LRDs. The red color of LRDs with gas-enshrouded AGN is produced by their prominent Balmer break, which is a sharp spectral feature, and the resulting color can be significantly impacted by the location of this break relative to the filters. This placement shifts significantly between our snapshots ($z=5$, 6, 7, 8), and as a result the criterion can accept or reject physically similar sources depending on the snapshot in which they are observed.
Omitting the criterion makes our selection slightly more permissive than a strict application of the UNCOVER criteria, but does not admit any of the contaminants that the criterion targets. In effect this increases the number of sources that are identified as LRDs (by $\sim20\%$), but provides a more consistent population that is still comparable to observed LRD populations based on their properties.

All sources are evaluated using both the Red1 and Red2 selection criteria. For presentation purposes, we assign each source a \redcolorLabel, and \bluecolorLabel\ based on their redshift. For sources below $z=6.5$, we assume they are more likely to be selected by the Red1 criteria, and thus assign their \redcolorLabel\ to be F200W-F356W, and their \bluecolorLabel\ to be F115W-F150W. For sources above $z=6.5$, we assume the Red2 criteria, assigning a \redcolorLabel\ of F277W-F444W, and their \bluecolorLabel\ of F150W-F200W. 

We present these selection criteria for the entire population of \ambra\ sources in figure \ref{fig:LRD_criteria}. The top row shows the observable properties of the sources, where all the AGN are assumed to be in a gas-enshrouded phase. As we discuss in sections \ref{subsec:AGN_SED} and \ref{subsec:LRD_lifecycle}, the duty cycle of this gas-enshrouded phase is likely much less than 1.0, and relies upon processes that occur well below the resolution of cosmological simulations. As such, these results represent all of the potential LRDs that could be observed, if their AGN are in a gas-enshrouded phase. To contextualize the impact of the gas-enshrouded model, we show the LRD population found in \ambra\ when applying a standard AGN model to all the sources, which is shown in the bottom row.

There are a large number of dim LRDs (F444W > 26.0) when assuming a gas-enshrouded AGN, and effectively none when assuming standard AGN emission. In contrast, the population of bright LRDs is fairly consistent across both models. This indicates the importance of an intrinsically red AGN spectrum to the appearance of dimmer LRDs. This is consistent with our expectations, as the reddening mechanisms such as dust attenuation are not as impactful in lower mass galaxies.

We compare the black hole luminosity function of our simulated population of LRDs in \ambra\ to those of observed LRD populations in figure \ref{fig:BH_LF}. 
\ambra\ is closest to reproducing the observed luminosity function at $z\sim5$--$6$, where the number of dim gas-enshrouded LRDs present in \ambra\ is very high. At these redshifts the gas-enshrouded LRDs in \ambra\ are comparable to the observations in the faintest bin ($\rm L_{bol} \sim 10^{43} ~erg~s^{-1}$), and fall $\sim 0.5$--$1$~dex below observations at $\rm L_{bol} \gtrsim 10^{44} ~erg~s^{-1}$. By $z\sim7$--$8$ the LRD luminosity function of \ambra\ falls $\sim2$ dex below the observations. While \ambra\ does not fully reproduce the observed LRD population, especially at higher redshifts, we do find a very large population of LRDs ($\sim 8,000$), which allows us to perform analysis of their properties, and the specifics of their LRD phases (see sections \ref{subsec:BH_gal_props} and \ref{subsec:LRD_lifecycle}).

The dotted lines in figure \ref{fig:BH_LF} show the effect of adding the unresolved AGN variability described in section \ref{subsec:AGN_SED}, with $\sigma_{\rm bol} = \AGNstoch$ dex. This scatter allows black holes whose luminosity in the simulation falls just below what is required to appear as an LRD to do so during their brighter phases, and it raises the LRD luminosity function at $\rm L_{bol} \sim 10^{44}~erg~s^{-1}$ by $\sim1$ dex at $z\sim5$--$6$ and by up to $\sim1.5$ dex at $z\sim7$--$8$, while leaving the bright end ($\rm L_{bol} \gtrsim 10^{45}~erg~s^{-1}$) essentially unchanged. With this variability included, and a unity duty cycle on the gas-enshrouded phase, the LRD luminosity function of \ambra\ is comparable to the observed number densities of \citet{Greene_2025} and \citet{Umeda_2026} at $z\sim5$--$6$, and the deficit at $z\sim7$--$8$ is reduced from $\sim2$ dex to $\lesssim1$ dex. The size of the high redshift deficit is therefore sensitive to how variability below the resolution of the simulation is treated, and a substantial part of it may reflect the instantaneous accretion rates available in the simulation. Across all redshifts, and regardless of the variability model, the distribution of the LRDs in \ambra\ is more closely aligned with the dimmer luminosity distribution proposed in \citet{Greene_2025} and \citet{Umeda_2026}, as the number of very bright AGN in the simulation, regardless of whether they reside in an LRD or not, is lower than some of the proposed observational luminosity functions.

\begin{figure}
    \centering
    \includegraphics[width=1.0\columnwidth]{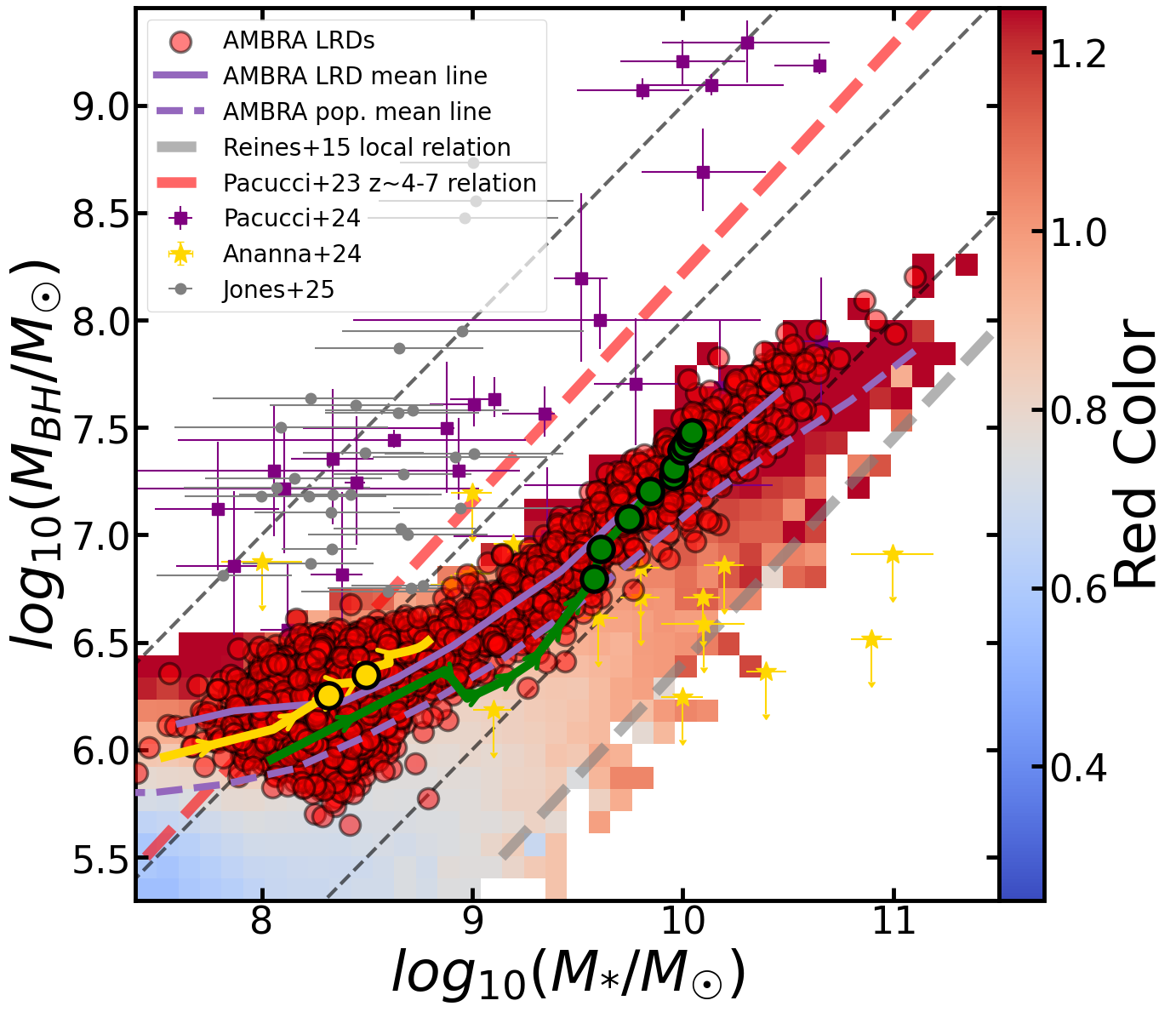}
    \caption{The $\rm M_{BH}-M_{\ast}$ relation for the black holes and their host galaxies in the \ambra\ simulation. \RedColorText. 
    \icontext. The purple square dataset is comprised of AGN fits from \citet{Maiolino_2024_JADES, Harikane_2023, Ubler_2023, Stone_2024, Furtak_2024, Yue_2024_eiger} compiled in \citet{Pacucci_Loeb_2024}. The gold star dataset was produced in \citet{Ananna_2024} from the X-ray observations of some LRDs, and the grey circles are the LRD sample of \citet{Jones_2025}. We also include the local relation per \citet{Reines_2015} and an inferred $z\sim 4$--$7$ relation from \citet{Pacucci_2023}. \tracktext. \meanlines.
    }
    \label{fig:MBH_Mstar}
\end{figure}

In addition to observations, and our \ambra\ results, we also include the mock LRD populations from \astrid\ \citep{LaChance_2025} and \brahma\ \citep{BRAHMA_LRDs} in magenta and green respectively. As \ambra\ is nearly identical to \astrid\ aside from the black hole seeding prescription, the significantly larger population of LRDs found in \ambra\ is indicative of the importance of this process to the production of LRDs. The comparison with \brahma\ is less direct, as there are many more differences between the two simulations. However, one important difference is that this \brahma\ realization uses a black hole repositioning scheme, whereas \ambra\ uses dynamical friction, which gives a more physical black hole evolution. The repositioning scheme results in faster mass assembly via mergers in the \brahma\ volume, which elevates the number of LRDs that are identified, but the black holes which power those LRDs have similar bolometric luminosities to those we find in \ambra.

In figure \ref{fig:BH_MF} we compare the overall black hole population, and LRD black hole population of \ambra\ to the observed broad-line AGN (BLAGN) population at similar redshifts.
This further reinforces the role of the gas-enshrouded AGN phase in the production of LRDs with lower mass, dimmer black holes. The black hole mass function also highlights the clearest difference from our previous \brahma\ results: the \brahma\ population has significantly more high mass black holes than we find in \ambra. This elevated black hole mass in \brahma\ results in the density of LRDs with  $\rm log_{10}(M_{BH} / \msun)\geq7$ eclipsing the total observed BLAGN population in that same mass range \citep{Taylor_2025b}. In contrast, the \ambra\ LRDs house black holes that are roughly 1 order of magnitude less massive than those in our previous \brahma\ results, so the number density of black holes in \ambra\ LRDs is generally well below the observed BLAGN population ($\sim 1$ dex compared to \citealt{Taylor_2025b}). The dotted line in figure \ref{fig:BH_MF} shows that the addition of unresolved AGN variability increases the LRD population almost entirely through low mass black holes: the number density of LRDs with $\rm M_{BH} \lesssim 10^{6}\msun$ increases by close to 1 dex, while the mass function is largely unchanged above $\rm 10^{6.5}\msun$. This is expected, as the higher mass black holes are already luminous enough to appear as LRDs and are unaffected by the additional scatter, whereas the far more numerous low mass black holes that sit just below the luminosity required for a source to appear as an LRD (see section \ref{subsec:BH_gal_props}) can be scattered above it. Even with this increase, the LRD black hole mass function of \ambra\ does not exceed the observed BLAGN mass function.

\begin{figure*}
    \centering
    \includegraphics[width=2.0\columnwidth]{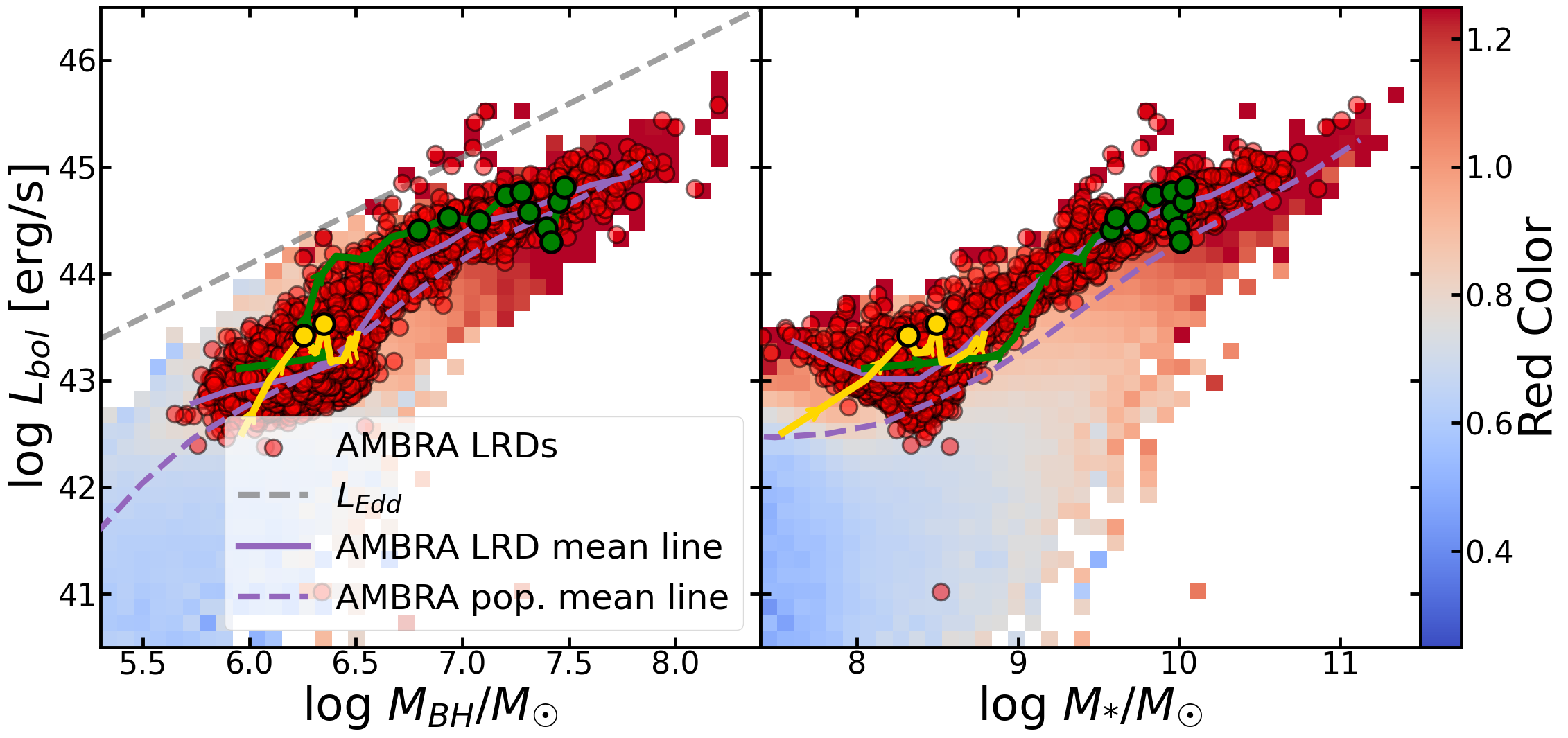}
    \caption{The $\rm L_{bol}-M_{BH}$ and $\rm L_{bol}-M_{\ast}$ relations for the black holes and their host galaxies in the \ambra\ simulation. \RedColorText. 
    The dashed grey line in the left panel indicates the Eddington luminosity as a function of black hole mass. \icontext. \tracktext. \meanlines.
    }
    \label{fig:Lbol_MBH_Mstar}
\end{figure*}

In figure \ref{fig:LRD_z_trend} we compare the overall redshift evolution of the LRD population in \ambra\ to our previous works, other theoretical works, and the observed LRD population. The observed LRD population slowly increases from $z\sim 9$ to $z\sim 5$ and then falls off towards lower redshifts \citep{Kokorev_2024, Akins_2025, Kocevski_2025, Zhuang_2026, Greene_2025}. The LRD population in \ambra\ is more steeply peaked at redshift 5, with a faster decline towards higher redshift. This decline is exaggerated by a dip at $z=7$. This dip is due primarily to the alignment of the Balmer break with the JWST NIRCam filters at $z=7$, as all of our observations occur in a single snapshot at that time. Exactly at $z=7$ the Balmer break falls just inside the F277W filter, leading to the measured F277W-F444W colors being decreased relative to physically similar galaxies at $z\sim7.25$.

The dotted line shows the LRD population of \ambra\ with the unresolved AGN variability of section \ref{subsec:AGN_SED} included. This raises the number density of LRDs by a factor of $\sim2$ at $z=5$--$6$, and $\sim10$ at $z=7$--$8$, which flattens the decline between $z=5$ and $z=8$ from $\gtrsim2$ dex to $\lesssim1.5$ dex. With this variability, the shape of the redshift evolution of the LRD population in \ambra\ is more similar to the observed redshift evolutions \citep{Kokorev_2024, Kocevski_2025, Akins_2025, Zhuang_2026,  Greene_2025}, although it still shows a steeper falloff towards high redshift than those observations. 

It is also very comparable to the other theoretical predictions shown in figure \ref{fig:LRD_z_trend}. \citet{Jeon_2025} analyze the A-SLOTH semi-analytic models (SAMs) with a variety of black hole seeding models, and find that a heavier seed mass both produces the peak in number density at $z\sim5$ and maintains a larger population towards high redshift than their lower seed mass model. This is consistent with our findings, as \ambra\ also features a heavy black hole seed mass, and once unresolved AGN variability is included, the LRD number density in \ambra\ tracks their heavy seed model across the entire redshift range we analyze in this work, despite the fact that \citet{Jeon_2025} select LRDs based on the properties of their black holes rather than on mock observables (see section \ref{subsec:comparison_w_other_works}). The LRD redshift evolution of the LUMINA simulation \citep{Lumina} also has a similar overall shape to the redshift evolution of LRDs in \ambra\ when modeled with the unresolved AGN variability. All of these theoretical results, including \ambra, show a steeper falloff towards high redshift ($\sim 1$--$1.5$ dex) than is seen in observations ($\lesssim 0.5$ dex). Successive refinements of the black hole modeling in these works (heavier seeds, unresolved variability) have already reduced this discrepancy, and it is possible that further iteration may remove it entirely. That said, this discrepancy may also be indicative of the presence of some other class(es) of objects \citep[see e.g.][]{Pacucci_2026_DCBH, Cenci_2025} that make up a larger portion of the LRD population at higher redshifts.

\subsection{Black hole and galaxy properties}
\label{subsec:BH_gal_props}

We examine the physical properties of the LRDs in \ambra\ and compare them with the rest of the galaxy population in order to determine which properties are associated with the LRD phase.

In figure \ref{fig:MBH_Mstar} we show the population of galaxies in \ambra\ on the $\rm M_{BH}-M_{\ast}$ plane, including those that appear as LRDs, and some JWST observational results. We find that \ambra\ does reproduce some of the overmassive black hole population, specifically for low mass host galaxies, but falls between the local relation, and a possible high redshift relation for host galaxies with stellar masses above $10^9 \msun$. The LRDs we identify in \ambra\ are concentrated near the upper portion of the $\rm M_{BH}-M_{\ast}$ envelope. Across the range of host galaxy stellar masses the LRDs house black holes that are on average twice as massive as the whole population of detectable galaxies. This aligns with the trend towards redder colors for host galaxies with higher mass black holes that is present for the whole population of galaxies in \ambra. While there is a correlation between higher mass black holes, and LRDs, there is a wide scatter across the LRD population, indicating that other properties play an important role in the production of an LRD.

In figure \ref{fig:Lbol_MBH_Mstar} we show the bolometric luminosities of black holes vs their own mass, and the stellar mass of their host galaxies. In the left panel, showing the $\rm L_{bol}-M_{BH}$ plane, we again see the correlation between black hole mass, and redder observed colors for the entire population of sources in \ambra. Somewhat surprisingly, there is only a small correlation between increasing black hole luminosity at a given mass, and redder observed colors. The LRD population is broadly constrained to black holes with bolometric luminosities of $10^{42.5}~\rm erg~s^{-1}$ or higher, but above that threshold there are objects with a wide range of Eddington ratios which appear as LRDs. This is highlighted by the track of the LRD shown in green. It first appears as an LRD when its black hole is near the top of the envelope ($\rm log_{10}(M_{BH}/M_\odot) \approx 6.8$ and $\rm log_{10}(L_{bol}/erg\,s^{-1}) \approx 44.4$, corresponding to $\lambda_{\rm Edd} \approx 0.3$), and it remains an LRD until $z=5$, when we reach the end of the simulation. This includes a period where the accretion of the black hole drops significantly, reaching a value of $\rm log_{10}(L_{bol}/erg\,s^{-1}) \approx 44.3$ at $\rm log_{10}(M_{BH}/M_\odot) \approx 7.4$ ($\lambda_{\rm Edd} \approx 0.06$), which is below the average for all black holes of similar mass. This is also evident from the LRD and population mean lines, which are quite similar. There is a preference for black holes with higher Eddington ratios at the low mass end ($\rm M_{BH} \lesssim 10^{6.25}\msun$) and at some intermediate masses ($\rm 10^{6.75}\msun \lesssim M_{BH} \lesssim 10^{7.5}\msun$), but the relatively small shift from the overall population, and the fact that it is inconsistent across the black hole mass range, indicate that the Eddington ratio is a weak predictor of whether or not a black hole will appear in an LRD.

In the right panel of figure \ref{fig:Lbol_MBH_Mstar} we show the $\rm L_{bol}-M_{\ast}$ plane, which shows a much stronger relationship between increased bolometric luminosity, and red color, with the LRD population residing near the top of the $\rm L_{bol}-M_{\ast}$ envelope. This indicates that the relative brightness of the AGN and the host galaxy is a stronger indicator of the potential for an object to appear as an LRD than the Eddington ratio of the black hole. This is in alignment with our expectations, especially when a gas-enshrouded AGN is present, as they have intrinsically red colors, so the more prominent the AGN is relative to its host galaxy, the redder it will appear. This is further emphasized when selecting LRDs, as a bright AGN can help produce the compact appearance of an LRD. The consistent gap between the means of the LRD population and the overall galaxy population (solid and dashed violet lines) further reinforces this point. On average, the AGN in LRDs are $\sim 3$ times as bright as the population average for a given host galaxy mass.

\begin{figure}
    \centering
    \includegraphics[width=1.0\columnwidth]{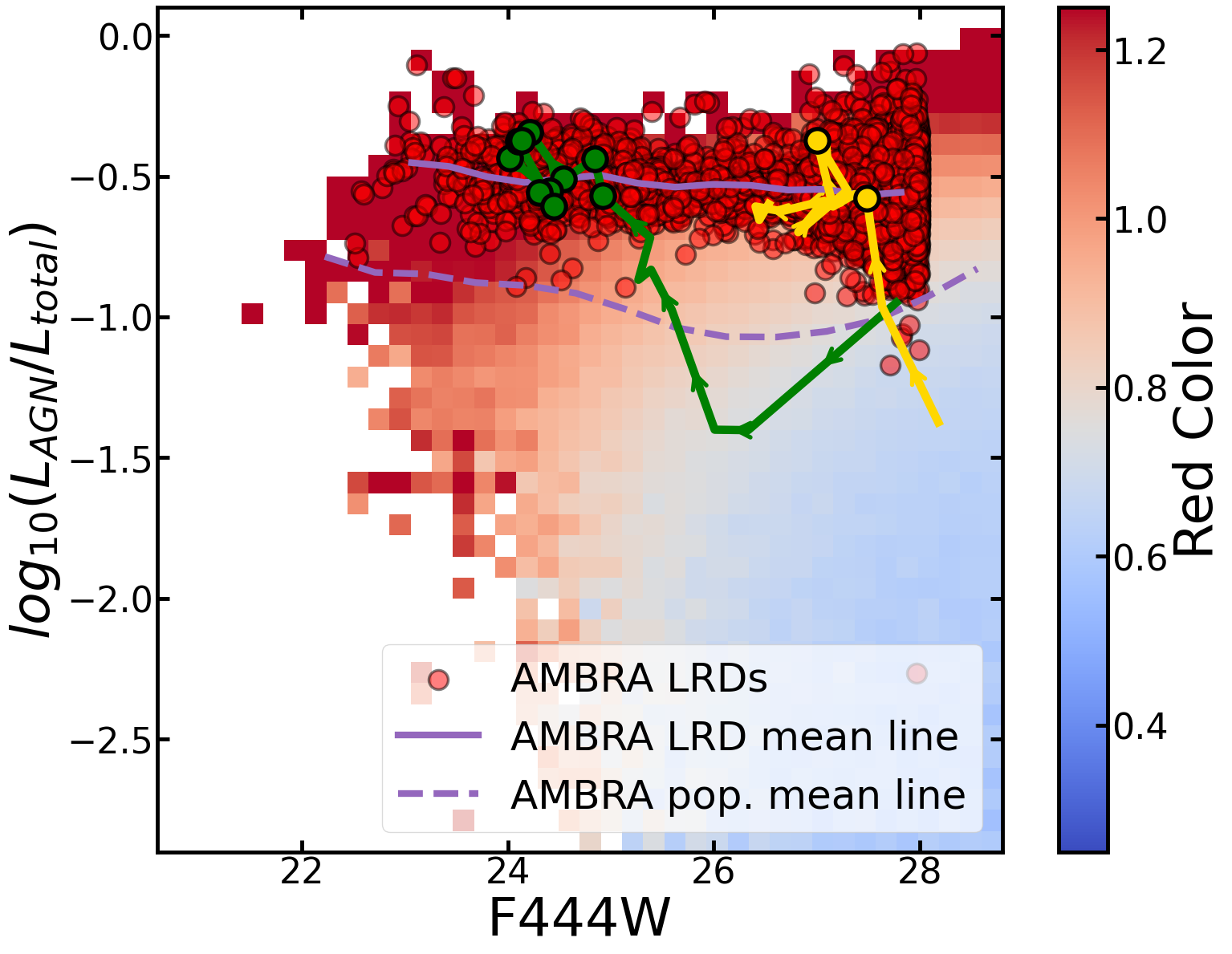}
    \caption{The AGN contribution to the F444W emission of the sources in \ambra. \RedColorText. \icontext. \tracktext. \meanlines.
    }
    \label{fig:F444W_AGN_contribution}
\end{figure}

We can see further evidence of this connection between AGN contribution, and redness in figure \ref{fig:F444W_AGN_contribution}. Here we see a strong correlation between the ratio of AGN to stellar emission in the F444W filter, and the color of the source. Nearly every single LRD identified in \ambra\ across all the snapshots we analyze has an AGN contribution above the population average for their observed magnitude ($10\%$), with the LRD average being roughly three times that of the population ($\sim30\%$). This factor of $\sim3$ is consistent between the AGN contribution results here, and the AGN luminosity vs host stellar mass results above.

\begin{figure}
    \centering
    \includegraphics[width=1.0\columnwidth]{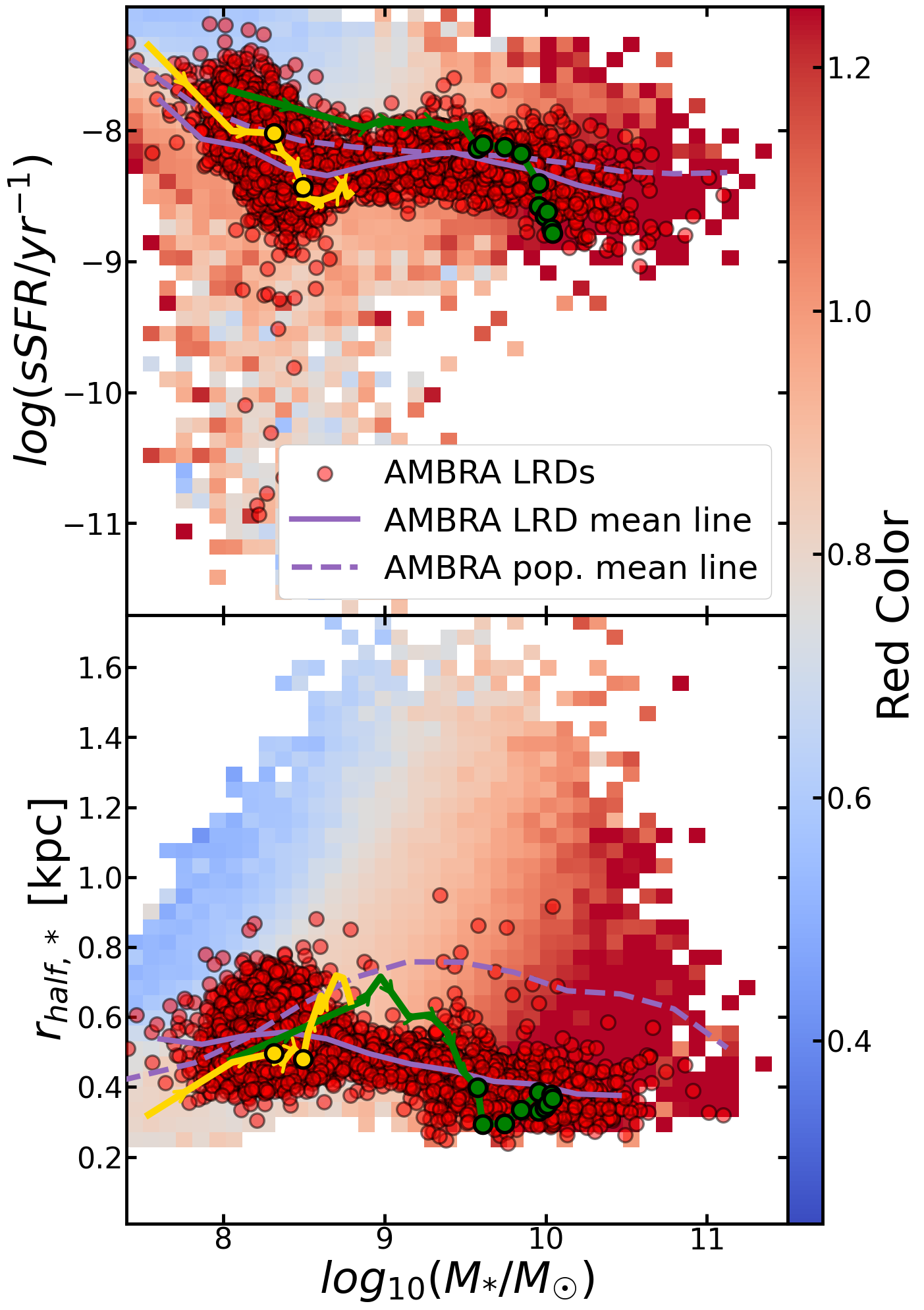}
    \caption{Galaxy specific star formation rate (sSFR) and stellar half-mass radius ($\rm r_{half,*}$) vs stellar mass shown in the top and bottom panels respectively. \RedColorText. \icontext. \tracktext. \meanlines.
    }
    \label{fig:sSFR_r_half}
\end{figure}

We also investigate the properties, and distribution of the host galaxies' stellar population, given they also account for a significant portion of the observed emission of these LRDs. We show the specific star formation rate (sSFR) and stellar half-mass radii of the galaxies in \ambra\ in figure \ref{fig:sSFR_r_half}. 

In the upper panel, showing the sSFR of the sources, we find a slight correlation between the star formation activity of a galaxy and its appearance as an LRD. The LRDs span nearly the full range of sSFR of the overall population, and the LRD mean line closely tracks the population mean line (solid and dashed violet lines respectively). The LRD mean is consistently less than or equal to the population mean, with the gap reaching a maximum of $\sim0.2$ dex at both $\rm log_{10}(M_{\ast}/\msun) \sim 8.5$ and $\sim 10.5$. This is consistent with the quenching that is produced by AGN feedback, which would be experienced at an above-average rate in these galaxies due to their brighter-than-average AGN.

In the lower panel showing the stellar half-mass radii of the sources, we see a correlation between smaller galaxies and redder colors that is present for all galaxy masses, in addition to the overall relationship between higher mass host galaxies and redder colors. The redder colors of more compact galaxies are likely the result of their higher dust column densities, which redden the emission of both the stars and AGN, in addition to them potentially hosting older stellar populations which appear redder than their younger counterparts. This contributes to the LRD population of \ambra\ residing in more dense host galaxies. In addition to the reddening effect of a more compact galaxy, these denser galaxies are also more likely to meet the compactness criterion required to be an LRD. 

This second effect is notably more important for higher mass objects. The LRDs with host galaxy masses above $10^{9.5} \msun$ have a mean half-mass radius of $\rm r_{half, \ast} \sim 0.4 ~kpc$ with very few above $\rm0.5 ~kpc$. On average, LRDs in this mass range are just over half the size of their non-LRD counterparts. In contrast, the LRDs with $M_{\ast} \leq 10^9 \msun$ have a larger mean half-mass radius of $\rm \sim 0.5 ~kpc$, with some as high as $0.8\rm ~kpc$. This indicates the selection for more compact galaxies is much stronger at higher masses. This is likely the result of two compounding factors. Physically, some of these higher mass LRDs may rely more strongly on dust attenuation for their red color. Observationally, a diffuse high mass host galaxy will still be detectable, and will cause the source to be too large to meet the compactness criterion. In contrast, a lower mass diffuse host galaxy is more difficult to separate from the background, and thus will have less of an impact on the measured compactness of the source. This can be seen in the population mean line (dashed violet line), which is significantly above the LRD population from $M_{\ast} \sim 10^{11} \msun$ down to $M_{\ast} \sim 10^{9} \msun$, where it begins to fall. It crosses below the LRD line by $M_{\ast} \sim 10^{8.25} \msun$. At these lower stellar masses, sources that are too diffuse are not distinguishable enough to be detectable, resulting in the observable population being more compact. The LRDs being less dense on average at the very low mass end is the result of LRDs housing brighter AGN, which aid in their detectability, allowing slightly more diffuse host galaxies to be detectable.

\subsection{The LRD Lifecycle}
\label{subsec:LRD_lifecycle}

In addition to the properties the LRDs have at the moment of observation, we analyze their histories to determine their lifetimes, and what may be responsible for the start and stop of the LRD phase. 

We include two tracks on figures \ref{fig:MBH_Mstar}--\ref{fig:sSFR_r_half} which show the time-series properties of the two LRDs shown in figure \ref{fig:histories_w_images}. In that figure we show the star formation rate, black hole accretion rate, compactness, and color histories of these two objects, along with a series of mock observations of them. The LRD in the top row (green track) is a higher mass LRD, which remains observable as an LRD for an extended period, including in the final snapshot of the simulation at $z=5$. The LRD in the bottom row (gold track) is a lower mass, dimmer LRD, which experiences two distinct, short LRD phases. These two objects are generally representative of the two broad categories of LRDs we find in \ambra. 

The first LRD is generally representative of the brighter LRDs in \ambra, which have higher mass black holes and host galaxies. These LRDs are generally less reliant on their AGN being gas-enshrouded, as we see a similar number of bright LRDs in both the standard AGN population, and the gas-enshrouded AGN population in figures \ref{fig:LRD_criteria} and \ref{fig:BH_LF}. This particular object begins appearing as an LRD around $z\sim6.5$, due to a combination of its AGN accretion rate increasing and its host galaxy becoming more compact. Physically, we can see in the upper history panel of the top row of figure \ref{fig:histories_w_images} that the accretion rate of the AGN is steadily increasing until the time when this object appears as an LRD. Similarly, while there is no signal of this in the star formation history, we can see in the bottom panel of figure \ref{fig:sSFR_r_half} that the host galaxy is becoming more and more dense until it reaches the LRD phase. Observationally, we see this as the AGN contribution to the F444W band increasing (see figure \ref{fig:F444W_AGN_contribution}), and the compactness dropping below 1.7. Once it reaches this point, we see variations in some of its properties, including a temporary decrease in the accretion (and thus luminosity) of the black hole, but it maintains its LRD appearance throughout these variations. 

The second is fairly representative of the dimmer LRDs, which have lower mass host galaxies. These LRDs are fully dependent on their AGN being gas-enshrouded in order to produce the red color that is integral to the appearance of an LRD. This is evident from the almost complete absence of LRDs with $\rm F444W \gtrsim 26.0$ from the standard AGN row in figure \ref{fig:LRD_criteria}, and the significant gap between the gas-enshrouded and standard AGN populations at the dim end of the black hole luminosity functions in figure \ref{fig:BH_LF}. This particular object undergoes two LRD phases, around $z\sim6.75$, and $z\sim6.0$. Both of these periods are short-lived, and directly correlated with periods where its AGN is bright. This is evident in figure \ref{fig:Lbol_MBH_Mstar}, as the two LRD phases (marked with gold circles) occur at moments where the AGN $\rm L_{bol}$ is spiking relative to the stellar mass. 

\begin{figure*}
    \centering
    \includegraphics[width=2.0\columnwidth]{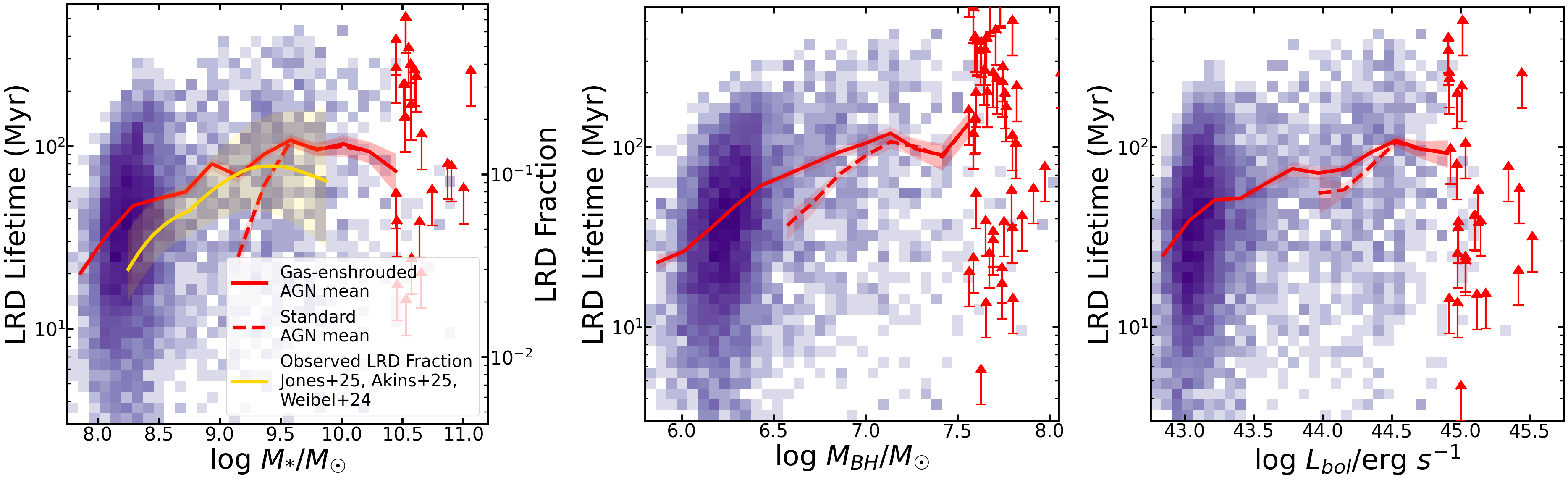}
    \caption{The length of the LRD window vs multiple physical properties of the LRDs. From left to right these are the stellar mass of the LRD host galaxy, the mass of the central black hole, and the bolometric luminosity of the black hole. The underlying 2D histogram shows the distribution of these values for all LRD windows we identify in \ambra, and the red line shows the mean of the LRD window length, while the shaded region is the uncertainty on that mean. We also include dashed lines which are the mean LRD lifetime for the LRDs identified with the standard AGN emission model. We truncate our mean lines when the number of sources in a bin drops below 10. We only include LRDs which conclude their LRD phase before $z=5$, as the total duration of the LRD phases that are still ongoing at $z=5$ is unknown. These objects make up a significant portion of the highest mass, and brightest LRDs, so we show the $z=5$ LRDs which fall above the cutoff for each of our mean lines as upward red arrows, as their LRD phase has not yet concluded. In the left panel we also show, as the gold line with its scale on the right-hand axis, the ``Observed LRD Fraction'' --- the fraction of observed galaxies that are LRDs as a function of stellar mass, constructed from the LRD populations of \citet{Akins_2025} and \citet{Jones_2025} and the galaxy stellar mass function of \citet{Weibel_2024} as described in section \ref{subsec:LRD_lifecycle} --- with the shaded region around it indicating its uncertainty.
    }
    \label{fig:LRD_window_len}
\end{figure*}

We analyze the entire population of LRD phases in figure \ref{fig:LRD_window_len}. We see that the vast majority of LRDs in \ambra\ have lifetimes between 3 and 300 Myr, with the higher mass, brighter LRDs having longer lifetimes ($\sim 100$~Myr) on average than lower mass, dimmer LRDs ($\sim 30$~Myr). This timescale is similar to both the Salpeter lifetime ($\sim45$~Myr; \citealt{Salpeter_1964}) and the typical quasar lifetime ($\sim 10$--$100$~Myr depending on obscuration; \citealt{Hopkins_2005}), which is expected given the AGN-centric nature of the LRDs we find in \ambra.

Notably, we do not include objects which are LRDs in the $z=5$ snapshot in the distribution for this analysis, as we do not know the full duration of the LRD phase. Many of the highest mass, and brightest LRDs in the simulation are active LRDs at $z=5$, so we present them as red arrows on figure \ref{fig:LRD_window_len} to provide a visual representation of where these high mass LRDs may reside. Additionally, for any object which requires a gas-enshrouded AGN to appear as an LRD, its lifetime presented here should be considered in conjunction with the duration of the gas-enshrouded phase. Depending on the source of the gas-enshrouded phase, it may only last $\lesssim 1$ Myr \citep{Inayoshi_2016, Takeo_2020, Shi_2023}, or up to tens of Myr \citep{Coughlin_2024, Begelman_2026, Sun2026}. In addition to the gas-enshrouded AGN lifetimes represented by the solid red line, we also present the standard AGN LRD lifetimes via the dashed red line. Here we can see that the two populations are nearly identical for the bright LRDs (those with high stellar mass, black hole mass, and black hole luminosity), supporting our conclusion that these LRDs are not dependent on the details of their AGN environment, and are instead primarily constrained by the compactness of their host galaxy. 

We also compare our galaxy mass-LRD lifetime trend with the fraction of observed galaxies that are LRDs at each galaxy mass, which we present as the gold line in figure \ref{fig:LRD_window_len}. 
While the two quantities are not directly analogous, they should share the same shape, with a normalization set by the duration of the period considered. In this case, we are considering the timeframe from $z=10$ to $z=5$ for our LRD lifetimes, which lasts $\sim 700$~Myr, so a mean LRD lifetime of 70 Myr would correlate with an ``Observed LRD Fraction'' of 10\%. We find this observed LRD fraction in the following manner. First, we find the LRD stellar mass function. We use the LRD population published in \citet{Akins_2025} as the normalization for the total number of LRDs, and the LRD population in \citet{Jones_2025} to determine the shape of the LRD stellar mass function. Then we divide this by the total galaxy stellar mass function (galaxy catalog from \citealt{Weibel_2024}) to determine the ``Observed LRD Fraction''.

We find that the shape of the ``Observed LRD Fraction'' curve is very comparable to the LRD lifetime curve of \ambra, and the ``Observed LRD Fraction'' curve is consistently below our LRD lifetime curve. This suggests our LRD lifetimes are consistent with the observed LRD and galaxy populations, with the vertical offset occurring for two reasons. First, some galaxies never appear as LRDs --- which would effectively count as lifetimes of 0.0 Myr, decreasing the mean LRD lifetime we find in \ambra. Second, some of the LRD lifetimes we find may be shortened by a transition out of the gas-enshrouded phase (as discussed above), which would also shift the mean LRD lifetime of \ambra.

\section{Discussion}
\label{sec:discussion}

\subsection{Comparison with other theoretical works}
\label{subsec:comparison_w_other_works}

The nature and evolution of LRDs have been explored in other recent theoretical works. These include other cosmological simulations like LUMINA \citep{Lumina} and MELIORA \citep{Cenci_2025}, and semi-analytic models like those presented in \citet{Jeon_2025}. These works provide important context for our results here, especially when considering the differences in both the underlying physical models, and the LRD formation scenarios that are explored in these works. 

One of the primary distinctions between our work and these other theoretical analyses is the method for identification of LRDs. As discussed in sections \ref{subsec:Obs_pipeline} and \ref{subsec:LRD_identification}, we create full mock observations of the sources in \ambra\ and use observables from those mock observations to determine which sources are LRDs, and which are not. The other current theoretical works identify LRDs or potential LRDs via the physical properties of the black holes present in their simulation or SAM. The combination of which LRD selection method is used, and the underlying physical models that are implemented in each of these simulations and SAMs, determines which LRD formation pathways are being explored in each work.

In the MELIORA simulation, \citet{Cenci_2025} identify all recently formed black holes as potential LRDs. They do so because their black hole seeding model is emulating DCBH formation conditions via e.g. the inclusion of a criterion on the Lyman--Werner flux. This facilitates the analysis of the population of objects which could be LRDs in the DCBH formation pathway scenario. They identify a population which undergoes a bright phase with bolometric luminosities $\gtrsim10^{43}~\rm erg~s^{-1}$ in a brief period ($\lesssim200~\rm Myr$) after their formation. This aligns well with proposed theoretical models of LRDs as recent DCBHs, such as \citet{Pacucci_2026_DCBH}, although the findings of \citet{Cenci_2025} support a UV-detectable host galaxy, unlike \citet{Pacucci_2026_DCBH}.

Our analysis in this work does not identify many of the young black hole LRDs that are explored in \citet{Cenci_2025}, as few of the black holes in \ambra\ reach luminosities of $\gtrsim10^{43}~\rm erg~s^{-1}$ in the $\sim 200 \rm ~Myr$ after their formation. This may be due to the lack of a Lyman--Werner criterion in black hole seeding, which allows for seeds to form earlier in \ambra, resulting in their seed environment being less dense, and thus less likely to facilitate the high accretion rates necessary to produce luminosities of $\gtrsim10^{43}~\rm erg~s^{-1}$ in young, lower mass black holes.

In the LUMINA simulation, \citet{Lumina} take an approach similar to \citet{Cenci_2025}, but apply an upper mass cut of $10\times\rm M_{seed}$ (to select ``young'' black holes) and a luminosity selection, and classify the objects which meet these criteria as LRDs. This has a similar focus on the recently formed black hole population as \citet{Cenci_2025}, but with the addition of a luminosity criterion in order to identify an LRD population, rather than the population of LRD candidates. Additionally, there is no DCBH-specific criterion in the black hole seeding prescription of LUMINA, remaining agnostic to the exact black hole formation pathway, and focusing on an empirical LRD description. 

With the black hole seed mass of \ambra\ covering a range of $\sim4.4\times10^{4}$--$4.4\times10^{5}\,M_{\odot}$, the upper limit of black hole masses which would be comparable to those found in LUMINA is $4.4\times10^{6}\msun$. A large number of the LRDs we identify do contain black holes which fall below this mass and have bolometric luminosities $\gtrsim10^{43}~\rm erg~s^{-1}$. That said, many more objects meet those criteria in \ambra\ but do not produce mock observables that meet the LRD criteria. This suggests that either the mock observation pipeline we employ here may not be properly reproducing the spectra of these objects, or the number of non-LRD objects that meet these criteria is higher than predicted. We intend to further explore this question in the future by applying the full LRD analysis method of \citet{Lumina} to \ambra\ and comparing the resulting LRD population with the one we find here, and those found in other works.

In their analysis of the A-SLOTH SAMs, \citet{Jeon_2025} take a different approach, focusing on higher mass black holes. In their heavy seed model, this is any black hole with $\rm M_{BH} \gtrsim 10^6 \msun$, and in their light, Pop III, seed model they apply the same mass threshold, along with a criterion that requires the black holes to be accreting at or above the Eddington rate. They find that the lighter seed models overpredict the density of LRDs, while the heavy seed models are broadly consistent with the observed LRD population. While the black hole seed model employed in \ambra\ is more analogous to the heavy seed models of \citet{Jeon_2025}, reproducing their LRD selection method would significantly overpredict the LRD population (see figure \ref{fig:BH_MF}), due to the lack of a Lyman--Werner seeding criterion, which is included in the heavy seed models of \citet{Jeon_2025}. Due to this difference in physical modeling it is difficult to draw a direct comparison between our results and those of \citet{Jeon_2025}, but the suppression in the LRD population they see due to the more stringent black hole seeding model they employ may be similar in magnitude to the suppression in the LRD population from our observational LRD criterion. While the overall number density of the LRD population predicted by these two models is somewhat consistent, the underlying physical properties are likely very different, and future analysis of how those differences may appear in observational properties could provide insights into which theoretical LRD population better represents the observed population.

\subsection{LRDs as a heterogeneous population}
\label{subsec:heterogeneous}

While a single LRD formation pathway may be responsible for the entire observed LRD population, it is also possible that the population is heterogeneous in nature, despite having fairly consistent observables. As we discuss throughout this work, the LRDs we identify in \ambra\ may not represent the entire observed population of LRDs. Without additional AGN variability their number density falls off too sharply towards higher redshift ($z\sim 7$--$8$), and they are unable to reproduce the observed LRD population without the gas-enshrouded phase having a significantly higher duty cycle than is predicted. While the inclusion of unresolved AGN variability (section \ref{subsec:LRD_identification}) reduces both of these tensions, they are still present. These LRDs fall within the category of pre-existing sources which undergo an LRD phase due to a combination of their black hole and host galaxy properties. In contrast, the other theoretical works, including \citet{Cenci_2025} and \citet{Pacucci_2026_DCBH}, examine the category of LRDs which occur briefly after the formation of a new black hole. \citet{Cenci_2025} suggest these objects are more likely to have a flatter redshift evolution, and a strong falloff towards lower redshifts ($ z \lesssim 5.5$). A combination of these two formation pathways may best describe the observed LRD population, with the DCBH pathway being more prevalent at higher redshift, and the pre-existing source pathway contributing most strongly near the peak at $z\sim5$. 

This hypothesis will be testable in the future, as we further develop our understanding of the differences between the observables of these two categories of LRDs, and gain access to a larger number of LRD observations across different redshifts. Possible probes of this nature include measurements of the metallicity of the AGN environment and host galaxy, with early results in this area already being published \citep{Nikopoulos_2026}. A larger sample that extends to higher redshift could provide evidence for or against these alternative LRD formation channels, such as recent DCBHs, via the fraction of LRDs with pristine gas environments. Similarly, future observations of ``little blue dot'' companions of LRDs \citep{Baggen_2026} could provide further support for the DCBH origin of some LRDs, and a possible redshift evolution of the fraction of LRDs with such companions could provide insights into the relative number of different classes of LRDs.

\section{Conclusions}
\label{sec:Summary}

In this work we investigate the population of little red dots (LRDs) in the \ambra\ cosmological hydrodynamic simulation, which pairs the volume, resolution, and astrophysical models of \astrid\ with a lenient, gas-based black hole seeding prescription taken from the \brahma\ simulation suite. We generate mock JWST NIRCam observations of the galaxy and AGN populations at $z = 5$--$8$ using \texttt{SYNTHESIZER}, modeling the stellar emission and dust attenuation of each source together with \texttt{cloudy}-based AGN templates, and apply observational LRD selection criteria to the resulting photometry and morphologies. We do this for both a standard AGN SED and a gas-enshrouded AGN SED in which a dense, turbulent gas cloud imprints a strong Balmer break on the transmitted spectrum. Because \ambra\ differs from \astrid\ only in its seeding model, and improves on our previous \brahma\ analysis by providing a $\sim 1000\times$ larger volume, it allows us to isolate the role of seeding in the production of LRDs. Unsurprisingly, when compared with those previous works we find a significantly higher density of LRDs in \ambra\ than \astrid, but a lower density than we found in \brahma, because of the reduction in overmassive black holes. So far, none of these simulations can fully account for the observed population, but the LRD population we find in \ambra\ represents a class of the observed LRDs, and we analyze their properties in this work. Additionally, we construct time-series mock observations of every LRD from $z = 10$ to $z = 5$, which allows us to follow the observable and physical properties of these sources through their LRD phase and place limits on LRD lifetimes.

We find that a gas-enshrouded AGN is instrumental to the production of dim LRDs. With the standard AGN SED, \ambra\ yields essentially no LRDs fainter than $m_{\rm F444W} \sim 26$, whereas assigning a gas-enshrouded SED to every AGN produces a large population of such sources, while leaving the number of bright LRDs largely unchanged. Under the extreme assumption of a unity duty cycle for the gas-enshrouded phase, the LRD luminosity function of \ambra\ is closest to observations at $z \sim 5$--$6$, where it is within $\rm\sim1~dex$, but falls $\sim 2$~dex below them by $z \sim 7$--$8$, and the LRD number density in \ambra\ declines by roughly two orders of magnitude between $z = 5$ and $z = 8$, more steeply than the gentle decline ($\rm \lesssim 0.5 ~dex$) seen in observations. Including a log-normal scatter of $\sigma_{\rm bol} = \AGNstoch$~dex in the bolometric luminosities of the black holes, to represent AGN variability below the resolution of the simulation, raises the LRD number density primarily through low mass black holes, flattens this decline to $\sim 1.5$~dex, and brings the shape of the redshift evolution in \ambra\ closer to that of observations, and into agreement with the heavy seed A-SLOTH model of \citet{Jeon_2025}, and the LUMINA simulation \citep{Lumina}. This indicates that the type of LRDs we find in \ambra\ contributes most to the LRD population near its peak at $z\sim5$. The steeper decline at earlier times may be resolved by either further iteration on the modeling of the AGN, or the presence of some other class of objects that appear as LRDs, and represent a significant portion of the LRD population at $z \gtrsim 7$. 

All of the LRDs we identify have some common properties. They occupy the upper envelope of the $M_{\rm BH}$--$M_*$ relation, hosting black holes that are on average twice as massive as those in the general population of detectable galaxies at fixed stellar mass, and the strongest predictor of LRD appearance is the brightness of the AGN relative to its host galaxy: LRD AGN are $\sim 3$ times more luminous than average at fixed $M_*$, and on average contribute $\sim 30\%$ of the source's F444W emission. The Eddington ratio, by contrast, is only a weak predictor, with LRDs spanning a wide range of $\lambda_{\rm Edd}$ above a luminosity floor of $L_{\rm bol} \sim 10^{42.5}$~erg~s$^{-1}$. Beyond these shared traits, however, the LRDs in \ambra\ separate into two broad categories, distinguished by their mass and brightness, that differ in which additional properties are required for them to appear as LRDs, and consequently in how long they remain LRDs.

The first category consists of the brighter, higher-mass LRDs, which reside in host galaxies with $M_* \gtrsim 10^{9.5}\,\msun$. These sources appear in similar numbers under both the standard and gas-enshrouded AGN models, and so do not depend very strongly on a gas-enshrouded phase to be identified as LRDs. Instead, their red color and compact morphology arise from a combination of a bright AGN and a host galaxy that is compact. The selection for compact hosts is especially strong at these masses: their mean stellar half-mass radius is $\sim 0.4$~kpc, with very few above $0.5$~kpc. Because their LRD appearance is driven by properties that evolve on galactic timescales rather than by transient events, these sources tend to be longer-lived, with LRD phases lasting $\sim 100$~Myr on average, and they can persist as LRDs through fluctuations in their accretion rate. Many of the most massive and brightest LRDs in \ambra\ are still LRDs at $z = 5$, so the full duration of their lifetimes may be longer still.

The second category consists of the dimmer, lower-mass LRDs, with host galaxies of $M_* \lesssim 10^{9}\,\msun$. These sources are almost entirely absent when a standard AGN SED is adopted, and are therefore fully dependent on their AGN being gas-enshrouded in order to produce the Balmer break responsible for their red color. Their LRD phases coincide with brief spikes in black hole accretion relative to the host stellar mass, and as the bright, intrinsically red AGN dominates both the color and the compact appearance of the source, they are less reliant on the compactness of their host galaxy, with a larger mean half-mass radius of $\sim 0.5$~kpc and some as extended as $0.8$~kpc. Because they require elevated accretion, which is a highly variable property, especially in these lower mass galaxies, these LRDs are short-lived, with lifetimes of $\sim 30$~Myr. Notably, their true lifetimes are further limited by the duration of the gas-enshrouded phase itself, which may be as short as $\lesssim 1$~Myr or as long as tens of Myr depending on its origin. The lifetimes we measure for this category should therefore be regarded as upper limits.

Taken together, these results indicate that the class of objects we identify as LRDs in \ambra\ is responsible for some or all of the observed LRD population: pre-existing black hole--galaxy systems that pass through an LRD phase driven by their AGN brightness relative to their host, and, depending on the brightness of the source, either a gas-enshrouded phase of the AGN or a compact host galaxy. Such sources likely comprise a significant fraction of the LRD population near its peak at $z \sim 5$, but our results indicate they become rarer toward higher redshift. This deficit is partially resolved by implementing the AGN variability found in \citet{Lumina}, indicating that further iteration on the models used in this work may fully resolve this high redshift LRD deficit. If not, this would indicate a heterogeneous LRD population, in which a second formation pathway, such as the luminous phases of recently formed heavy seeds explored by \citet{Cenci_2025} and \citet{Pacucci_2026_DCBH}, dominates at $z \gtrsim 7$. Distinguishing between these classes of objects observationally, and applying a consistent set of selection criteria across simulations that capture each pathway, will be essential to determining what fraction of the LRD population each of them represents.

\section*{Acknowledgments}
AKB and PT acknowledge support from NSF-AST 2510738, NSF-AST 2346977, and the NSF-Simons AI Institute for Cosmic Origins which is supported by the National Science Foundation under Cooperative Agreement 2421782 and the Simons Foundation award MPS-AI-00010515.
TDM acknowledges funding from NASA ATP 80NSSC20K0519, NSF PHY-2020295, NASA ATP NNX17AK56G, NASA ATP 80NSSC18K101, and NASA Theory grant 80NSSC22K072.
TDM and YZ acknowledge the support from the NASA FINESST grant 80NSSC25K0318.
FP acknowledges support by the Black Hole Initiative at Harvard University.
LB acknowledges support from NSF award AST-2307171 and NASA award 80NSSC22K0808.
SB acknowledges funding from NSF AST-2509639.
\astrid\ and \ambra\ were run on the Frontera facility at the Texas Advanced Computing Center.


\bibliographystyle{mnras}
\bibliography{references} 

\end{document}